\documentclass[letterpaper]{article}
\usepackage[margin=1in]{geometry}
\usepackage[colorinlistoftodos]{todonotes}
\usepackage{authblk}
\usepackage{amsmath}
\usepackage{amssymb}
\usepackage{multicol}
\usepackage{subfigure}
\usepackage{booktabs,multirow}
\usepackage{xcolor}
\usepackage{tabularx}
\usepackage{siunitx}
\DeclareSIUnit[quantity-product = {}]{\poise}{P}
\DeclareSIUnit[quantity-product = {}]{\dyne}{dyne}
\usepackage{verbatim}
\usepackage[title]{appendix}

\newcolumntype{L}{>{\raggedright\arraybackslash}m{2cm}}
\newcolumntype{M}{>{\raggedright\arraybackslash}m{3.5cm}}
\DeclareMathOperator{\Tr}{tr}

\usepackage{hyperref}
\hypersetup{
    colorlinks=true,
    linkcolor=blue,
    filecolor=magenta,
    urlcolor=cyan,
    citecolor=blue,
    breaklinks=true
}
\newcommand*{\email}[1]{\href{mailto:#1}{\nolinkurl{#1}}}

\title{Comparative Assessment of Biomechanical Parameters in Subjects With Multiple Cerebral Aneurysms Using Fluid--Structure Interaction Simulations\footnote{Published as \textit{ASME J.\ Biomech.\ Eng.}\ \textbf{145} (2023) 051003. DOI: \href{https://dx.doi.org/10.1115/1.4056317}{10.1115/1.4056317}. }}
\author{
Tanmay C.\ Shidhore$^{1}$,
Aaron A.\ Cohen-Gadol$^{2}$,
Vitaliy L.\ Rayz$^{1,3,\dagger}$, and
Ivan C.\ Christov$^{1,\dagger}$\\
{\small\it $^{1}$School of Mechanical Engineering, Purdue University, West Lafayette, Indiana 47907}\\
{\small\it $^{2}$Department of Neurological Surgery, Indiana University School of Medicine, Indianapolis, Indiana 46202}\\
{\small\it $^{3}$Weldon School of Biomedical Engineering, Purdue University, West Lafayette, Indiana 47907}\\
{\small $\dagger$ Corresponding authors; \email{vrayz@purdue.edu}, \email{christov@purdue.edu}.}
}

\begin{document}

\maketitle

\begin{abstract}
Cerebral aneurysm progression is a result of a complex interplay of the biomechanical and clinical risk factors that drive aneurysmal growth and rupture. Subjects with multiple aneurysms are unique cases wherein clinical risk factors are expected to affect each aneurysm equally, thus allowing for disentangling the effect of biomechanical factors on aneurysmal growth. Towards this end, we performed a comparative computational fluid--structure interaction analysis of aneurysmal biomechanics in image-based models of stable and growing aneurysms in the same subjects, using the cardiovascular simulation platform SimVascular. We observed that areas exposed to low shear and the median peak systolic arterial wall displacement were higher by factors of 2 or more and 1.5, respectively, in growing aneurysms as compared to stable aneurysms. Furthermore, we defined a novel metric, the oscillatory stress index (OStI), that indicates locations of oscillating arterial wall stresses. We observed that growing aneurysms were characterized by regions of combined low wall shear and high OStI, which we hypothesize to be associated with regions of collagen degradation and remodeling. Such regions were either absent or below 5\% of the surface area in stable aneurysms. Our results lay the  groundwork for future studies in larger cohorts of subjects, to evaluate the statistical significance of these biomechanical parameters in cerebral aneurysm growth.
\end{abstract}

\section{Introduction}
\label{sec:Intro}

Cerebral or intracranial aneurysms are dilations in the walls of cerebral arteries, most commonly occurring in or near the Circle of Willis. Cerebral aneurysms are estimated to occur in 5-8\% of the population \cite{RDAV98}. Aneurysm rupture results in subarachnoid hemorrhage with a 50\% mortality rate \cite{S97}. However, the mechanisms driving growth and rupture of aneurysms are poorly understood, and clinicians are often faced with difficult decisions weighing potential complications of invasive treatment against the risk of aneurysm rupture. Therefore, there has been increased interest in subject-specific computational modeling of blood flow in cerebral aneurysms to augment medical imaging data with high-resolution hemodynamic information that could be used to assess the risk of aneurysm rupture and, eventually, provide clinicians with better rupture risk stratification tools. While quantifying the hemodynamic forces acting on the vessel wall endothelium through flow-only computational fluid dynamics (CFD) simulations can elucidate their effect on aneurysm progression, the assessment of aneurysm stability is incomplete without analyzing the wall deformation and the mechanical stresses within the vessel wall. Most modeling studies do not involve fluid-structure interaction (FSI) simulations, as standard-of-care imaging data lacks resolution to reliably detect subject-specific wall thickness and composition.

Nevertheless, previous computational studies have highlighted the importance of accounting for this flow-vessel wall interaction. Torii \textit{et al.}~\cite{TOKTT09} demonstrated a reduction in maximum wall shear stress (WSS) by as much as 20\% at locations of flow impingement in patient-specific FSI models of some aneurysm subjects as compared to WSS predictions from corresponding rigid-wall simulations. Takizawa \textit{et al.}~\cite{TBTC12} also compared WSS, oscillatory shear index (OSI), and arterial stress and stretch, computed from three-dimensional (3D) FSI simulations of ten subject-specific geometries, to their rigid wall counterparts at mean arterial pressure conditions. Specifically, Takizawa \textit{et al.}~\cite{TBTC12} found an overestimation of WSS and qualitative differences in OSI computed from the rigid-wall and FSI simulations. 

Aneurysm progression is a complex interplay between abnormal biomechanical factors and environmental or clinical risk factors (e.g.\ genetic predisposition, smoking status, family history, etc). Therefore, elucidating the impact of biomechanical parameters alone on aneurysm progression would require controlling these clinical risk factors. Subjects with multiple unruptured aneurysms are unique cases that allow us to control for these subject-specific confounding factors. We hypothesize that clinical risk factors mentioned previously would affect all aneurysms in the same subject equally. Therefore, any differences in aneurysm progression may be attributed to differences in biomechanical factors alone. Previous flow-only and FSI studies did not account for means to control clinical risk factors and moreover, did not analyze subjects with multiple aneurysms. In the present study, we address this knowledge gap in the literature by providing a framework to compare biomechanical parameters between stable and growing aneurysms in the same subject with an objective to reveal the influence of local flow and mechanics on aneurysm progression, and eventually rupture.

To this end, this paper is organized as follows. In Section~\ref{sec:Comp_modeling}, we describe the process of extracting the computational geometries of the aneurysm and estimating the vessel wall thickness using standard medical imaging data. In Section~\ref{sec:Num_framework}, we briefly present the numerical framework used by the open-source cardiovascular modeling tool SimVascular, in particular, the svFSI solver, which we used to run FSI simulations. Then, Section~\ref{sec:Results} summarizes our main results. Specifically, we compare biomechanical parameters such as wall shear stress (WSS), oscillatory shear index (OSI), arterial wall displacement, and the orientation of principal stresses in the arterial wall between stable and growing aneurysms in the same subject. Finally, Section~\ref{sec:Conc} discusses limitations, future work, and open questions. 

\section{Anatomical and Computational Modeling}
\label{sec:Comp_modeling}

\subsection{Subject Data}

\begin{figure}[ht]
    \centering
    \includegraphics[width=0.75\textwidth]{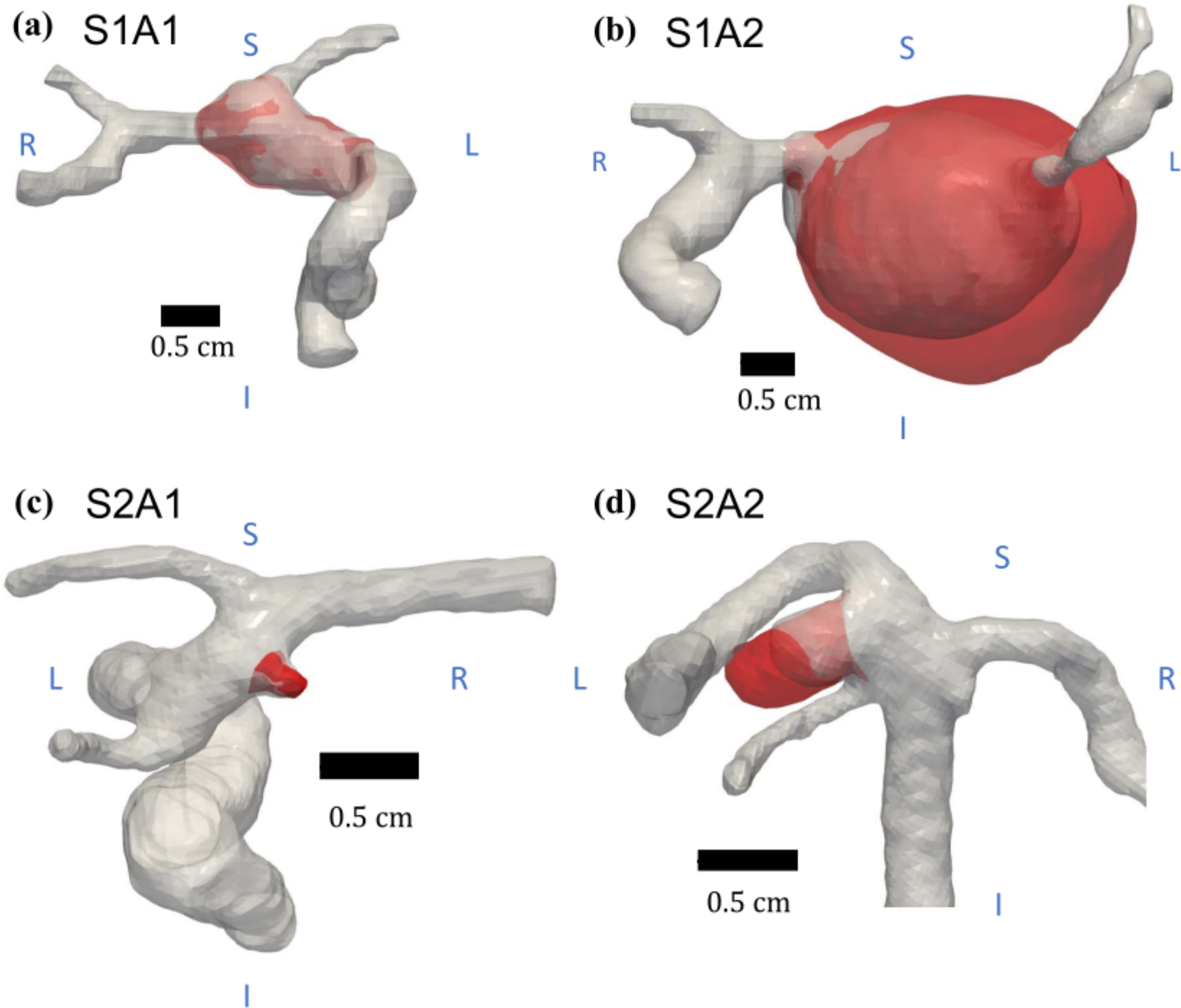}
    \caption{(a)-(d): Subject-specific baseline (solid) and follow-up (transparent) geometries of stable and growing aneurysms. 
    Abbreviations used -- R: Right, L: Left, S: Superior, I: Inferior.}
    \label{fig:Geometry}
\end{figure}

Under a protocol approved by an Institutional Review Board, retrospective CT angiography images for two subjects, each with two unruptured cerebral aneurysms were obtained from the Indiana University School of Medicine. Images were available at baseline and at one year post baseline  (hereafter referred to as `follow-up'). Subject 1 (hereafter referred to as `S1') presented with a fusiform aneurysm of the right internal carotid artery, which was stable (labeled `S1A1'), and a fusiform aneurysm of the left middle cerebral artery, which grew in size between baseline and follow-up imaging (labeled as `S1A2'). Similarly, subject 2 (hereafter referred to as `S2') presented with two saccular aneurysms, one of the proximal right posterior communicating artery that remained stable (labeled as `S2A1') and one on the left superior cerebellar artery, which grew (labeled as `S2A2'). Figure~\ref{fig:Geometry}(a)-(d) shows the solid baseline geometries, along with the superposed transparent follow-up model, visually demonstrating the extent of growth. Table~\ref{tb:Subject_characteristics} reports the subject characteristics, along with the percentage of increase in volume for the growing aneurysms between baseline and follow-up studies.

\begin{table}[ht!]
\caption{Subject and aneurysm characteristics.}
\begin{center}
\begin{tabular}{lllLLlL}
\toprule
& & \multicolumn{3}{c}{Growing Aneurysm} & \multicolumn{2}{c}{Stable Aneurysm}\\\cline{3-7}
Subject & Age, Sex & Location & Initial diam.\ (mm) & Volumetric growth (\%) & Location & Initial diam.\ (mm)\\[2mm]
\midrule
S1 & 62, Male & L MCA & 22 & 65 & R ICA& 9 \\
S2 & 67, Female & L SCA& 5 & 10.5 & R PCommA& 3 \\
\bottomrule
\end{tabular}
\end{center}
\label{tb:Subject_characteristics}
{\noindent Abbreviations used -- R: Right, L: Left, MCA: middle cerebral artery inferior bifurcation, ICA: internal carotid artery, PCommA: posterior communicating artery, SCA: superior cerebellar artery.}
\end{table}

\subsection{Computational Geometries}

\subsubsection{Flow Domain Geometry}

We segmented the CT angiographic images for each aneurysm using the open-source 3D medical image analysis tool ITK-SNAP \cite{YPHSHGG06}. The domain of interest consisted of the aneurysm, along with the parent/feeding vessel and the immediate outlet branches. The geometries were subsequently smoothed and cleaned up using the commercial computer-aided design (CAD) tool Geomagic\textsuperscript{\textregistered} Design X to eliminate any segmentation artifacts such as sharp edges and bumps. The accuracy of the segmentations, as well as identification and removal of any segmentation artifacts, was ensured by consulting with our clinical collaborators throughout the process of generating the computational geometries.

\subsubsection{Wall Thickness Estimation and Solid Domain Geometry}
\label{sss:WT_estimation}

\begin{figure}
    \centering
    \includegraphics[width=0.75\textwidth]{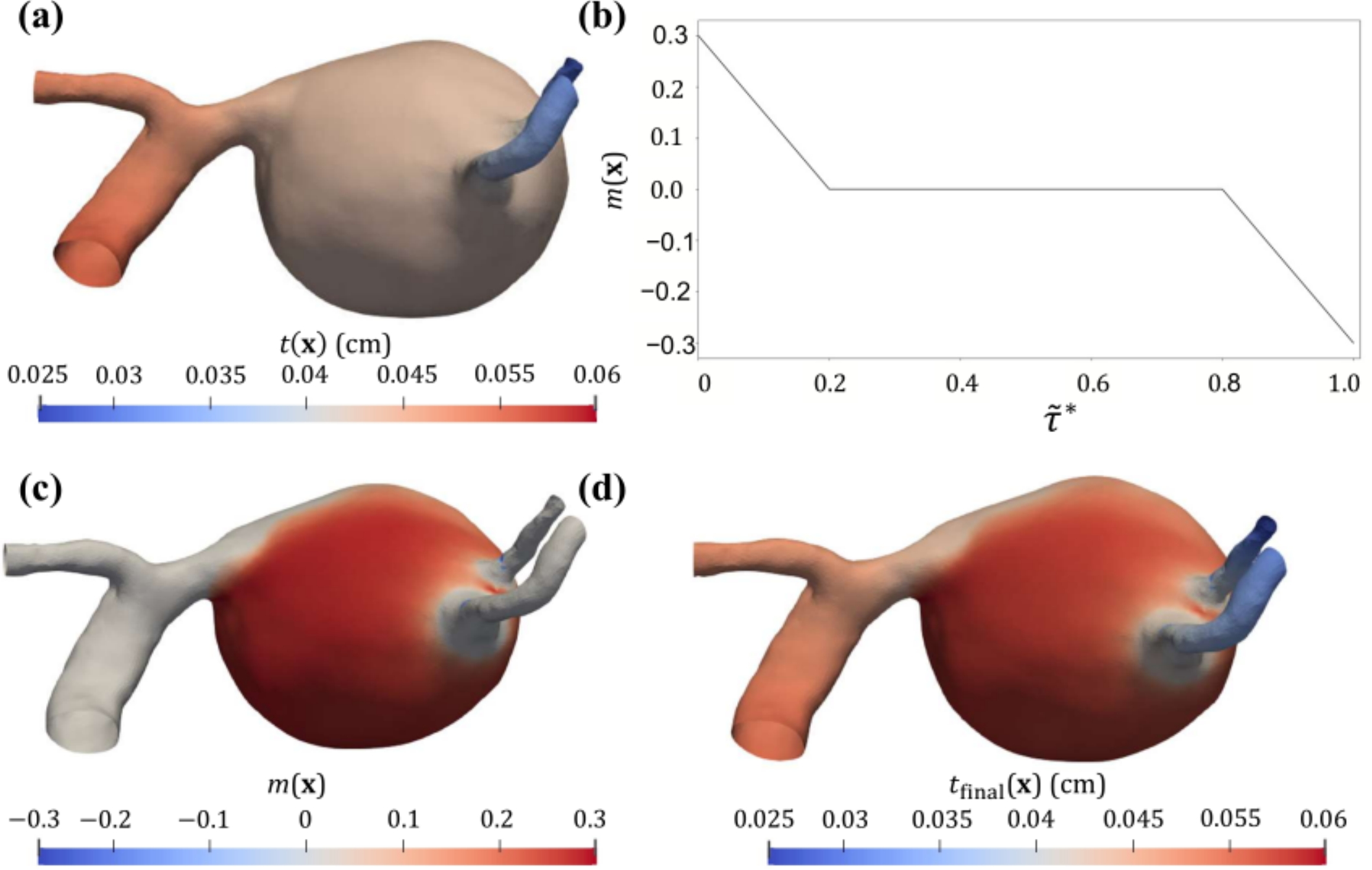}
    \caption{Methodology to estimate nonuniform arterial wall thickness from subject-specific aneurysm geometry. (a) Baseline nonuniform wall thickness $t(\mathbf{x})$ estimated using Equation \eqref{Eqn:Laplace}. (b) Graphical representation of the modulation function $m(\mathbf{x})$ from Equation \eqref{Eqn:Modulation} as a function of the normalized time-averaged wall shear stress $\tilde{\tau}^*$ from Equation \eqref{Eqn:WT_normalized_WSS}. (c) Wall thickness modulation as implemented for aneurysm S1A2 (see Figure \ref{fig:Geometry}(b)). (d) Final nonuniform wall thickness $t_{\rm final}(\mathbf{x})$ for aneurysm S1A2 obtained using Equation \eqref{Eqn:Wall_thickness}.}
    \label{fig:Wall_thickness}
\end{figure}

Accounting for variation in arterial wall thickness is crucial to obtaining physiologically realistic FSI simulation results \cite{CVSHTSBLP15,BKVVTLRZ16}. In particular, Vo\ss\ \textit{et al.}~\cite{VGHBWJPTJB16} found differences in wall shear stress distribution between FSI simulations conducted with constant and subject-specific wall thickness values. In the present study, the available CT angiography data lacks resolution to obtain reliable information on the vessel wall. Therefore, we implemented a workflow to estimate the nonuniform wall thickness, accounting for the effect of heterogeneous distribution of hemodynamic forces on the vessel wall. We estimated the nonuniform baseline arterial wall thickness by solving a Laplace equation over the luminal surface, as originally proposed by Bazilevs \textit{et al.}~\cite{BHBSM09}. Specifically, the vessel wall thickness $t(\mathbf{x})$ at location $\mathbf{x}$ along the luminal surface (i.e.\ the inner wall $\partial\Omega_{\rm inner}$) was found as the solution to:
\begin{equation}
    \nabla^2 t(\mathbf{x}) = 0.
    \label{Eqn:Laplace}
\end{equation}
The domain for the Laplace equation \eqref{Eqn:Laplace} is the inner wall $\partial\Omega_{\rm inner}$, and boundary conditions are prescribed on the sets of curves forming the inlets and outlets of the computational geometry.  Figure \ref{fig:Wall_thickness}(a) shows an example of the baseline wall thickness estimate obtained for S1A2. 

Additionally, we use the formulation proposed by Cebral \textit{et al.}~\cite{CVSHTSBLP15} to modulate $t(\mathbf{x})$ over the aneurysm using time-averaged wall shear stress data obtained from a rigid-wall pulsatile flow simulation. However, unlike \cite{CVSHTSBLP15}, where the modulation was determined based on the absolute time-averaged wall shear stress, we use the normalized time-averaged wall-shear stress $\tilde{\tau}^*$ given by
\begin{equation}
    \tilde{\tau}^*(\mathbf{x}) = \frac{\tilde{\tau}(\mathbf{x})}{\tilde{\tau}_{\rm max}}.
    \label{Eqn:WT_normalized_WSS}
\end{equation}
Here, $\tilde{\tau}(\mathbf{x})$ is the time-averaged wall shear stress magnitude at a given location $\mathbf{x}$ on the aneurysmal surface and $\tilde{\tau}_{\rm max}$ is the spatial maximum time-averaged wall shear stress magnitude over the aneurysm, both of which, as mentioned earlier, are obtained via pulsatile rigid-wall flow simulations. 

Furthermore, we consider a different (as compared to \cite{CVSHTSBLP15}) modulation function:
\begin{equation}
    m(\mathbf{x}) =
    \begin{cases}
        0.3 - 1.5\tilde{\tau}^*, &\quad\tilde{\tau}^*(\mathbf{x}) < 0.2,\\
        1.2 - 1.5\tilde{\tau}^*, &\quad\tilde{\tau}^*(\mathbf{x}) > 0.8,\\
        0, &\quad\text{otherwise}.
    \end{cases}
    \label{Eqn:Modulation}
\end{equation}
Based on this definition of $m(\mathbf{x})$, it can be seen that regions of low and high wall shear stress (i.e.\ for which $\tilde{\tau}^* < 0.2$ and $\tilde{\tau}^* > 0.8$) are thickened or thinned up to a maximum of 30\% of the baseline value \cite{CVSHTSBLP15}. Meanwhile, regions of moderate wall shear stress (i.e.\ for which $0.8 \ge \tilde{\tau}^* \ge 0.2$) are left unchanged. The final wall thickness $t_{\rm final}(\mathbf{x})$ is calculated as:
\begin{equation}
    t_{\rm final}(\mathbf{x}) = \big[1+m(\mathbf{x})\big]t(\mathbf{x}).
    \label{Eqn:Wall_thickness}
\end{equation}
Figure~\ref{fig:Wall_thickness}(b)-(d) shows plots of $m(\mathbf{x})$ along with an example of the above modulation function and the final wall thickness for aneurysm S1A2. We would like to emphasize that we do not claim that the actual subject-specific wall thickness of each aneurysm is given by Equation~\eqref{Eqn:Wall_thickness}. Nevertheless, the proposed methodology is a practical methodology that enables accounting for the local effects of aneurysmal wall shear on wall thickness in the subsequent FSI simulations.

\subsubsection{Mesh Generation}

An unstructured mesh with tetrahedral elements was generated using ANSYS Workbench, for both the fluid and solid domains. Accounting for the requirements of the svFSI solver, it was ensured that both meshes were conformal (i.e.\ had a node-to-node matching) at the fluid-solid interface. The meshes were then imported into the svFSI simulation workflow via a processing pipeline developed in-house. A mesh resolution of $0.03$~\si{\centi\meter} was chosen for the fluid domain. To ensure the accuracy of shear stress computations, we refined the local mesh near the fluid-solid interface in the flow domain up to a constant thickness of $0.03$~\si{\centi\meter}. A grid independence analysis was performed on one of the aneurysm geometries to ensure that the computational results were independent of the core and near-wall mesh refinement resolutions (see Appendix A of the Supplemental Material for further details). To satisfy the requirement of conformity of the fluid and solid meshes at the fluid-solid interface, a surface mesh resolution of $0.03$~\si{\centi\meter} was also chosen for the solid mesh. Furthermore, it was ensured that the solid mesh contained two layers of tetrahedral elements along the wall thickness.

\section{Governing Equations and Numerical Framework}
\label{sec:Num_framework}

The open-source cardiovascular modeling platform SimVascular \cite{UWMLMS17,LUWMSM18}, in particular, the svFSI solver \cite{svFSI} was used to perform pulsatile 3D numerical simulations. Both the flow and structural problems are solved in svFSI using a finite element formulation. An arbitrary Lagrangian--Eulerian (ALE) approach is employed for the FSI coupling. We refer the reader to \cite{BVSSCMCFFMF20,VLXKHM17} for additional details. We only discuss the governing equations being solved; further details of the numerical schemes and their implementation may be found in \cite{BCHZ08,EBM13,THZ98}.

\subsection{Fluid Domain}

Blood was assumed to be a Newtonian fluid. This modeling assumption is commonly used in medium-sized arteries, such as those in the Circle of Willis \cite{TOKTT07}. Furthermore, the flow of blood was modeled as an incompressible flow. Identical blood density and viscosity values ($\rho_f = 1.06~\si{\gram/\centi\meter\cubed}$, $\mu_f=4~\si{\centi\poise}$) were used for all subjects' simulations. These values were taken from previous literature \cite{BVSSCMCFFMF20}. The incompressible Navier--Stokes equations representing the conservation of linear momentum and mass, respectively, were used to model the flow in the fluid domain. In the ALE formulation, they can be written as follows:
\begin{subequations}\begin{align}
        \rho_f\big[\dot{\mathbf{v}} + (\mathbf{v} - \hat{\mathbf{v}})\cdot\nabla\mathbf{v} \big] &= -\nabla p + \mu_f\nabla^2 \mathbf{v}, \\
        \nabla\cdot\mathbf{v} &= 0.
        \label{Eqn:Fluid_flow}  
\end{align}\end{subequations}
Here, $\hat{\mathbf{v}}$ is the grid velocity of the flow domain, $\dot{\mathbf{v}}$ is the ALE time derivative of the flow velocity $\mathbf{v}$, $\nabla$ represents the Eulerian spatial gradient operator, and $\mu_f$ and $\rho_f$ are the dynamic viscosity and density of blood, respectively.

\subsection{Structural Domain}

The elastic arterial vessel wall was modeled as a nearly incompressible isotropic neo-Hookean material. A Young's modulus $E=10^7~\si{\dyne/\centi\meter\squared}$ and a Poisson ratio $\nu=0.49$ were assumed for both subjects. Again, these values were obtained from previous literature \cite{TOKTT09}. In the simulations, we employed a large-deformation formulation for the elasticity problem, which involves a mapping between the coordinates in the current configuration $\mathbf{x}$ and the reference configuration $\mathbf{X}$. Using the large-deformation formulation, the balance of linear momentum in the material gives the dynamic equations of the motion. Neglecting external forces, these take the form:
\begin{equation}
     \rho_s\ddot{\mathbf{u}} + \nabla_{\mathbf{X}}\cdot \big( \mathbf{F} \mathbf{S} \big) = \boldsymbol{0}.
     \label{Eqn:Structural_dynamics}
\end{equation}
Here, $\rho_s$ is the density of the solid, $\mathbf{u} = \mathbf{x}-\mathbf{X}$ is the displacement, $\ddot{\mathbf{u}}$ is the acceleration, and $\mathbf{F} = \nabla_{\mathbf{X}}\mathbf{x} = \mathbf{I} + \nabla_{\mathbf{X}} \mathbf{u}$ is the deformation gradient, where $\nabla_{\mathbf{X}}$ represents the gradient operator with respect to the reference configuration, and $\mathbf{I}$ is the identity tensor. 

In Equation~\eqref{Eqn:Structural_dynamics}, the second Piola-Kirchhoff stress tensor $\mathbf{S}$ is determined from the constitutive relation proposed in \cite{SH99}. Specifically, it is derived from a strain-energy density function $\psi$ as:
\begin{subequations} \begin{align}
    \mathbf{S} = 2\frac{\partial \psi}{\partial \mathbf{C}} &= \mu_s J^{-2/3}\left(\mathbf{I} -\frac{1}{3}(\Tr{\mathbf{C}})\mathbf{C}^{-1}\right) + \frac{1}{2}\kappa\left(J^2-1 \right)\mathbf{C}^{-1},\\
    \psi(\mathbf{C},J) &= \frac{1}{2}\mu_s\left(J^{-2/3}\Tr{\mathbf{C}} - 3\right) + \frac{1}{2}\kappa\left[\frac{1}{2}(J^2-1) - \ln{J}\right],
    \label{Eqn:Constitutive_model}
\end{align}\end{subequations}
where $J=\det\mathbf{F}$ is the Jacobian, and $\mathbf{C}=\mathbf{F}^\top\mathbf{F}$ is the right Cauchy-Green deformation tensor. Note that, in Equation~\eqref{Eqn:Constitutive_model}, the material parameters $\mu_s$ and $\kappa$, respectively the shear and bulk moduli of the solid, are used instead of $E$ and $\nu$. For neo-Hookean materials, these four material parameters are related in the same way as in linear elasticity \cite{SH99}: $\mu_s = {E}/[2(1+\nu)]$, $\kappa = {E}/[3(1-2\nu)]$.

\subsection{Tissue Pre-Stress}
\label{ss:Pre_stress}

As pointed out in several previous studies \cite{TMWPMCWT08,TT14,TSSC08,TTBC11}, the vascular geometry obtained from imaging data is not stress-free but rather under continuous mechanical loading due to blood pressure. Therefore, appropriate initialization of the loading state of the geometry is necessary to obtain accurate vessel wall displacements. In this analysis, we used the methodology proposed by Hsu and Bazilevs \cite{HB11}. The second Piola-Kirchhoff stress tensor $\mathbf{S}$ in Equation \eqref{Eqn:Structural_dynamics} is augmented by an additional pre-stress tensor $\mathbf{S}_0$. That is, $\mathbf{S}$ is replaced by $\mathbf{S} + \mathbf{S}_0$ in Equation \eqref{Eqn:Structural_dynamics}. Here, $\mathbf{S}_0$ is defined such that it is in equilibrium with the incoming blood flow's tractions at cardiac-cycle averaged conditions.
In this study, the flow traction data was obtained for each aneurysm geometry via separate pulsatile rigid-walled flow simulations. Using these simulations as an input, the pre-stress tensor $\mathbf{S}_0$ was estimated for all aneurysmal geometries and prescribed as an initial stress state in subsequent FSI simulations. Additional details on the implementation of the pre-stress estimation methodology in svFSI can be found in \cite{BVSSCMCFFMF20}.

\subsection{Boundary Conditions}

\subsubsection{Fluid Domain Boundary Conditions}

\begin{figure}
    \centering
    \includegraphics[width=0.75\textwidth]{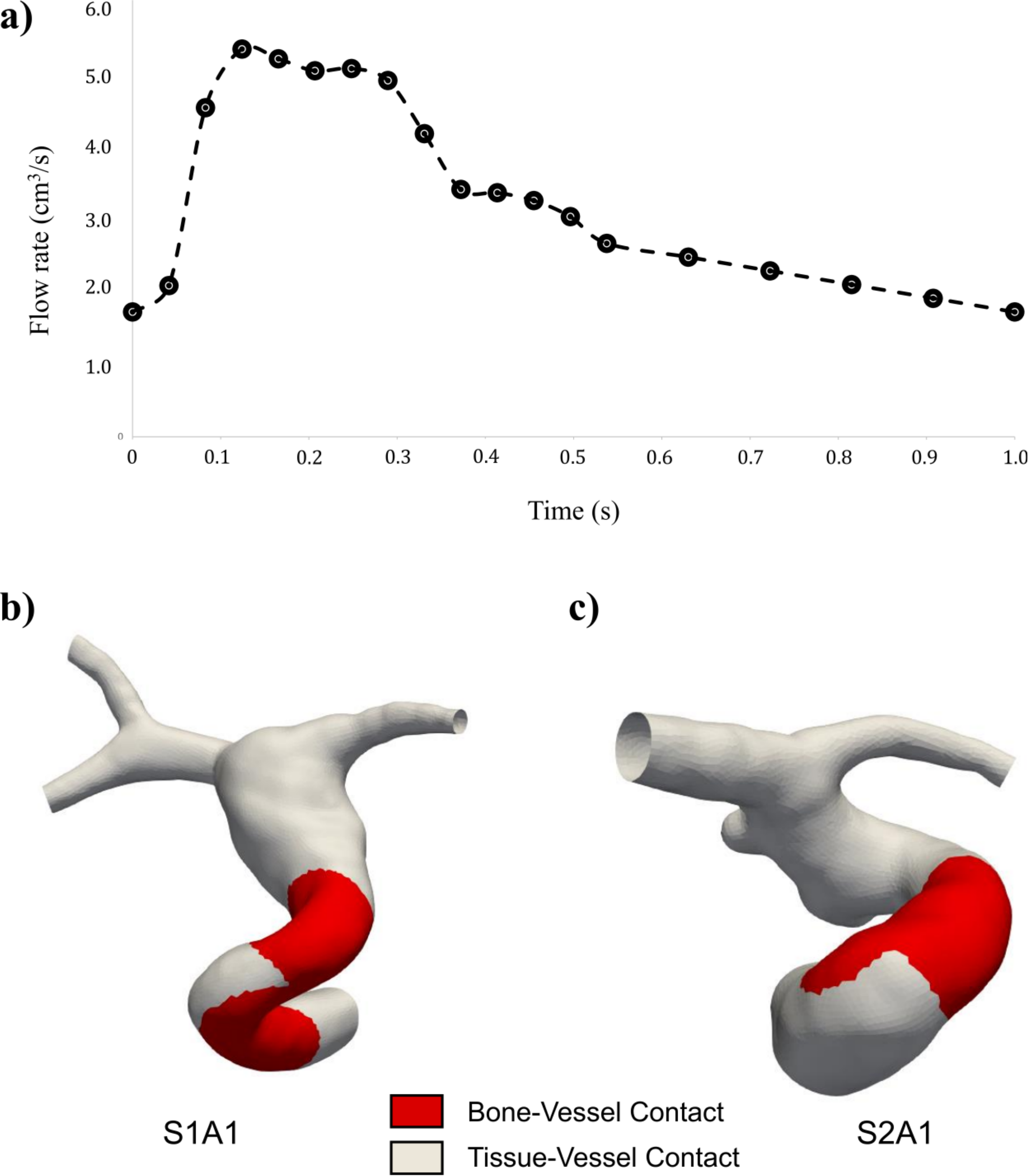}
    \caption{(a) Inlet pulsatile flow rate (in $\si{\centi\meter\cubed/\sec}$) obtained from MR measurements in the middle cerebral artery of a healthy volunteer. Depending on the inlet vessel for each aneurysm, the flow rate values were scaled to match population-averaged centerline velocities at peak systole data from \cite{MF15}. (b) and (c) Partitions on the outer vessel wall for aneurysms S1A1 and S2A1 (see Figures \ref{fig:Geometry}(a) and (c)) used to impose nonuniform spring constant values, depending on the nature of contact for the traction condition in Equation \eqref{Eqn:Spring-dashpot}.}
    \label{fig:BCs}
\end{figure}

A pulsatile flow profile acquired from MR (magnetic resonance) measurements in the middle cerebral artery of a healthy volunteer was prescribed at the inlet, as shown in Figure \ref{fig:BCs}(a). To determine the spatial velocity profile to be specified, the Womersley number, ${\rm Wo} = R\sqrt{{2\pi f \rho_f}/{\mu_f}}$, was estimated based on the radius of the parent/feeding vessel $R$ and the cardiac frequency $f$ (beats per second). As the ${\rm Wo}$ values for all aneurysm geometries were found to be between $2$ and $4$, a parabolic (Poiseuille) flow profile was implemented at the inlet cross-section. To account for differences in the feeding artery in each aneurysmal geometry, the flow profile acquired from the health volunteer was scaled such that the centerline velocity at peak systole matched previously reported population-averaged PC-MRI (phase contrast magnetic resonance imaging)  measurements \cite{MF15}. The scaling factor $s$ was determined as
\begin{equation}
    s = \frac{\tilde{v}_{\rm ps, a}}{\tilde{v}_{\rm ps, e}},
    \label{Eqn:Flow_rate_scaling_factor}
\end{equation}
where, $\tilde{v}_{\rm{ps, a}}$ is the above-mentioned population-averaged PC-MRI measurement of centerline velocity at peak systole, and $\tilde{v}_{\rm ps, e}$ is calculated as
\begin{equation}
    \tilde{v}_{\rm ps, e} = 2\cdot\frac{Q_{\rm ps}}{A_{\rm i}}.
    \label{Eqn:Estimated_centreline_velocity}
\end{equation}

Here, $Q_{\rm ps}$ is the flow rate at peak systole in the healthy volunteer flow profile (see Figure \ref{fig:BCs}a) and $A_{\rm i}$ is the cross-sectional area at the inlet. Therefore, $\tilde{v}_{\rm ps, e}$ may be regarded as the centerline velocity in the case where the healthy volunteer inflow profile is being imposed directly at the inlet (without any scaling). It should be noted that the pre-factor of 2 in Equation \ref{Eqn:Estimated_centreline_velocity} arises due to the cross-sectional flow profile being parabolic, as explained earlier.

To model the effect of the downstream vasculature, a three-element RCR (or, `Windkessel') boundary condition was imposed at each of the outlets \cite{VFJT10}. The initial estimates of the total arterial resistance $R^0_{\rm total}$ and capacitance $C^0_{\rm total}$ were obtained as:
\begin{subequations}\begin{align}
    R^0_{\rm total} &= \frac{\tilde{P}-P_0}{\tilde{Q}}, \\
    C^0_{\rm total} &= \frac{Q_{s} - Q_{d}}{P_s - P_d}\Delta t.
\end{align}\end{subequations}
Here, $\tilde{P}$ and $\tilde{Q}$ are the time-averaged pressure and flow rate, respectively, over a single cardiac cycle,  while $P_s$, $Q_s$ and $P_d$, $Q_d$ are the systolic and diastolic pressures and flow rates, respectively; $P_0$ is the distal pressure, and $\Delta t$ is the time difference between $Q_s$ and $Q_d$.

Subsequently, as outlined in \cite{XAF14}, we tuned the distal pressure, total resistance, and capacitance in an iterative fashion such that both diastolic, $P_d$,  and pulse, $P_s - P_d$, pressures were within $10\%$ of the corresponding normal values (i.e.\ $80~\si{mmHg}$ and $40~\si{mmHg}$, respectively). We ran rigid-wall pulsatile flow simulations for six cardiac cycles, and the results from the fourth cardiac cycle were used in the fine-tuning process. We further distributed the resistance across each individual outlet using the allometric scaling law ($m=2.4$) used for cerebral vessels \cite{BSCCR21}:
\begin{equation}
     R_{{\rm out},\ell}=\frac{\sum_{k=1}^{p} \sqrt{A_k^m}}{\sqrt{A_\ell^m}}\cdot R_\mathrm{total}.
     \label{Eqn:Murray_law}
\end{equation}
Here, $R_\mathrm{total}$ is the net downstream resistance, $A_{\ell}$ is the cross-sectional area of the $\ell^\text{th}$ outlet, and $p$ is total the number of outlets. The capacitance of each individual outlet branch is calculated as proposed in \cite{XAF14}:
\begin{equation}
     C_{{\rm out},\ell}=\frac{R_{\ell}}{R_\mathrm{total}}\cdot C_\mathrm{total} .
     \label{Eqn:Cap_split}
\end{equation}
For each outlet branch, the ratio of the distal to proximal resistance was assumed to be 1:9 \cite{BVSSCMCFFMF20}. Tables \ref{tb:S1_RCR} and \ref{tb:S2_RCR} list the RCR parameters used for each aneurysm.

\begin{table}[ht]
\begin{center}
\begin{minipage}{\textwidth}\footnotesize
\caption{RCR parameter values at each outlet for subject S1.~\label{tb:S1_RCR}}
\begin{tabular*}{\textwidth}{@{\extracolsep{\fill}}llllll@{\extracolsep{\fill}}}
\toprule
Aneurysm & Outlet & {$R_p$} ($\times 10^3~\si{\dyne\sec\per\centi\meter\tothe{5}})$ & {$R_d$} ($\times 10^4~\si{\dyne\sec\per\centi\meter\tothe{5}})$ & {$C$} ($\times 10^{-7}~\si{\centi\meter\tothe{5}/\dyne})$ & {$P_0$} (\si{mmHg}) \\
\midrule
\multirow{3}{*}{S1A1} & R MCA inferior & 5.19&4.67&1.53&\multirow{3}{*}{70}\\
& R MCA superior & 5.48 & 4.93 & 1.45 & \\
& R ACA & 12.6 & 11.4 & 0.63 & \\
\midrule
\multirow{3}{*}{S1A2} & L MCA inferior & 3.32 & 2.99 & 11.6 & \multirow{3}{*}{70} \\
& L MCA superior & 9.27 & 8.34 & 4.14 & \\
& L ACA & 2.42 & 2.18 & 15.9 & \\
\bottomrule 
\end{tabular*}
\vskip 3mm
{\noindent Abbreviations used -- R: Right, L: Left, MCA inferior: middle cerebral artery inferior bifurcation, MCA superior: middle cerebral artery superior bifurcation, ACA: anterior cerebral artery, $R_p$: proximal resistance, $R_d$: distal resistance $C$: capacitance, $P_0$: distal pressure.}
\end{minipage}
\end{center}
\end{table}

\begin{table}[ht]
\begin{center}
\begin{minipage}{\textwidth}\footnotesize
\caption{RCR parameter values at each outlet for subject S2.~\label{tb:S2_RCR}}
\begin{tabular*}{\textwidth}{@{\extracolsep{\fill}}llllll@{\extracolsep{\fill}}}
\toprule
Aneurysm & Outlet & {$R_p$} ($\times 10^3~\si{\dyne\sec\per\centi\meter\tothe{5}})$ & {$R_d$} ($\times 10^4~\si{\dyne\sec\per\centi\meter\tothe{5}})$ & {$C$} ($\times 10^{-7}~\si{\centi\meter\tothe{5}/\dyne})$ & {$P_0$} (\si{mmHg}) \\
\midrule
\multirow{3}{*}{S2A1} & R MCA & 1.82&1.64&43.6&\multirow{3}{*}{70} \\
& R ACA & 7.32 & 6.59 & 10.9 & \\
& R PCommA & 10.2 & 91.5 & 7.83 & \\
\midrule
\multirow{3}{*}{S2A2} & R PCA & 6.47 & 5.83 & 12.5 & \multirow{3}{*}{70} \\
& L PCA & 6.71 & 60.4 & 12 & \\
& L SCA & 24.7 & 24.9 & 2.92 & \\
\bottomrule 
\end{tabular*}
\vskip 3mm
{\noindent Abbreviations used -- R: Right, L: Left, MCA: middle cerebral artery inferior bifurcation, ACA: anterior cerebral artery, PCommA: posterior communicating artery, SCA: superior cerebellar artery, PCA: Posterior cerebral artery, $R_p$: proximal resistance, $R_d$: distal resistance $C$: capacitance, $P_0$ = distal pressure.}
\end{minipage}
\end{center}
\end{table}

\subsubsection{Structural Domain Boundary Conditions}

For the solid caps at each flow outlet, a homogeneous Dirichlet boundary condition, $\mathbf{u} = \boldsymbol{0}$, was imposed. The most common boundary condition imposed on the outer surface of the vessel wall is a no-traction boundary condition (see \cite{BTT13} and the references therein). However, as reported by Moireau \textit{et al.}~\cite{MXAFCTG12}, this simplification, which neglects the effect of surrounding tissue, induces unphysiological motion patterns in the vessel wall. Here, we account for the effect of surrounding tissue by using the Robin (spring-dashpot) boundary condition proposed in \cite{MXAFCTG12}. Mathematically, this condition is expressed via the traction $\boldsymbol{\sigma}\cdot\mathbf{n}$, which is determined by the local displacement $\mathbf{u}$ and velocity $\dot{\mathbf{u}}$. Specifically, the condition imposed on the outer wall $\partial\Omega_{\rm outer}$ is
\begin{equation}
    \boldsymbol{\sigma}\cdot\mathbf{n} = -k\mathbf{u} - c\dot{\mathbf{u}} - p_0\mathbf{n} \quad\text{ on }\quad\partial\Omega_{\rm outer},
    \label{Eqn:Spring-dashpot}
\end{equation}
where $k$ is an elastic spring constant, $c$ is a viscous damping coefficient, $p_0$ is a constant pressure value that can represent the intracranial/intrathoracic pressure, and $\mathbf{n}$ is the local unit normal. This approach essentially models the tethering of the outer wall to a fictitious medium that provides support similar to a Kelvin-Voigt viscoelastic material. 

Following previous work by B{\"a}umler \textit{et al.}~\cite{BVSSCMCFFMF20}, we simplify the tissue support model to account for only linear elastic effects by setting $c=0$ and $p_0=0$,  identically. However, we prescribe a spatially nonuniform spring constant $k$ depending on the type of contact (i.e.\ bone/tissue) as shown in Figure~\ref{fig:BCs}(b) and (c). To maintain consistency, $k$ values were kept identical for the same type of contact across both subjects. Contact between vessel and bones, which is expected to exhibit smaller displacements, was modeled by prescribing a large value ($k=10^9~\si{dyne\per\centi\meter\cubed}$). 

For the other regions of the vessel, as explained by Moireau \textit{et al.}~\cite{MXAFCTG12}, the tissue support parameter(s) must be calibrated by matching wall displacement data from simulations to \textit{in vivo} imaging. Here, since the available data was in the form of static images with no information on vessel wall displacements, we performed simulations with three different values (i.e.\ $k=10^4$, $10^5$ and $10^6~\si{dyne\per\centi\meter\cubed}$), which were within range of previously used values for $k$ in literature \cite{MXAFCTG12,BVSSCMCFFMF20}. In Section \ref{sec:Results}, we report data for the case with $k=10^6~\si{dyne\per\centi\meter\cubed}$, which corresponded to the case resulting in peak systolic displacements closest to the physiologically realistic values, based on the expert opinion of our clinical collaborators.

\subsection{Initial Conditions}

In order to ensure that the initial transients associated with ``flow start-up'' were eliminated in the least number of cardiac cycles, appropriate initial conditions were imposed in our FSI simulations.
The flow rate corresponding to cardiac-cycle average conditions was chosen as the starting point ($t=0$) of the cardiac cycle. This choice ensured consistency with the pre-stress calculations from Section \ref{ss:Pre_stress} and enabled the prescription of zero displacement and pre-stress for the initial displacement and stress state, respectively, of the elastic solid. The initial conditions in the flow domain were taken to be the cardiac-cycle averaged flow solution (velocity and pressure), obtained from the previous  rigid-walled flow simulation, which was also used to generate $\tilde{\tau}^*$ (recall Section \ref{sss:WT_estimation}).

\section{Results and Discussion}
\label{sec:Results}

\begin{figure}
    \centering
    \includegraphics[width=0.75\textwidth]{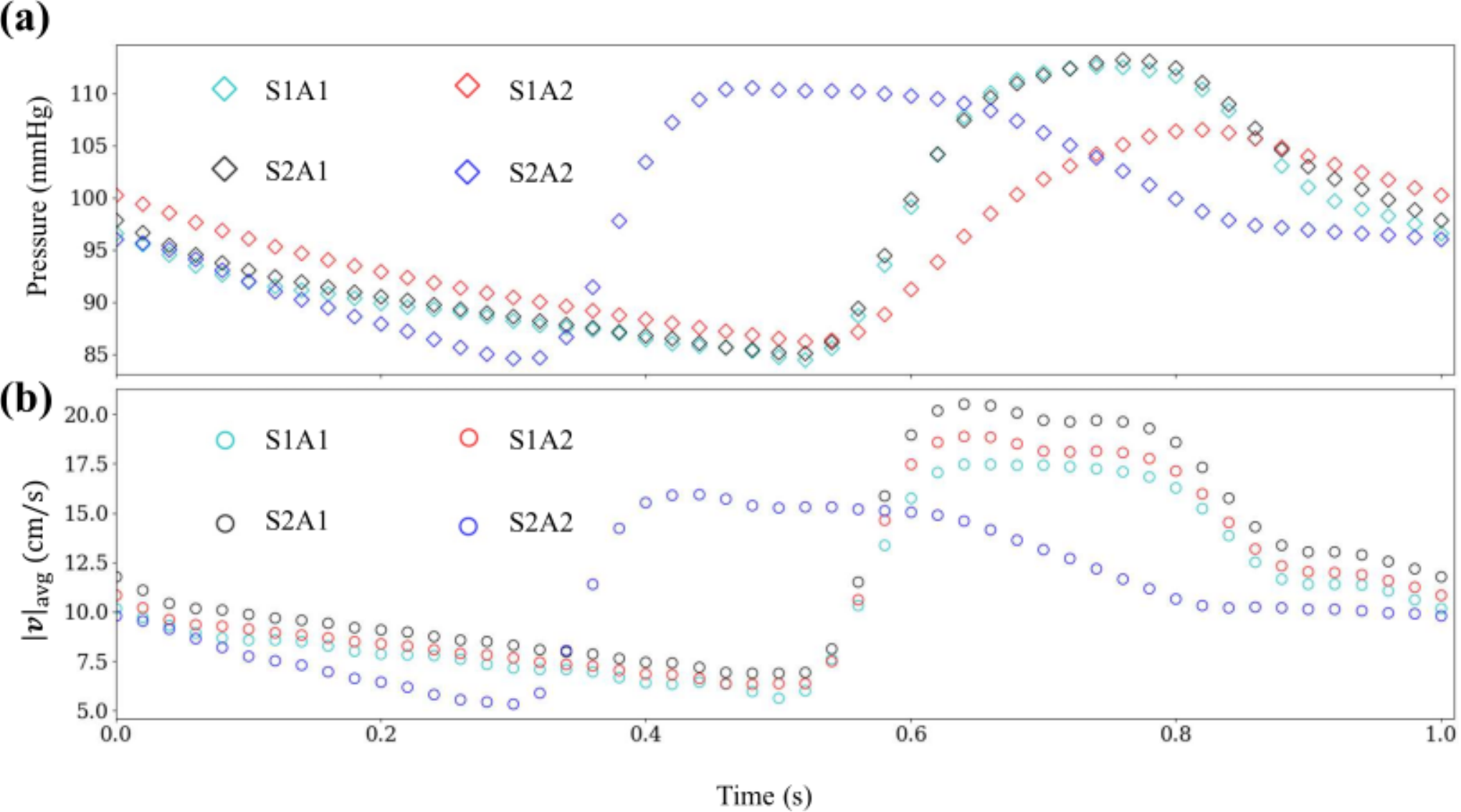}
    \caption{(a) Area-averaged pressure at the inlet of each aneurysm for the last cardiac cycle. (b) Area-averaged inflow velocity magnitude $|v|_{\rm avg}$ at the inlet of each aneurysm for the last cardiac cycle. In both panels, the color of each symbol represents a particular aneurysm geometry.}
    \label{fig:Inflow_AA_p_and_v}
\end{figure}

FSI simulations were run for four cardiac cycles and periodicity was found to be achieved after the 2\textsuperscript{nd} cardiac cycle. We report results from the last cardiac cycle. Figure \ref{fig:Inflow_AA_p_and_v} shows the area-averaged pressure and velocity magnitude waveforms at the inlet of each aneurysm. 

\begin{figure}
    \centering
    \includegraphics[width=0.75\textwidth]{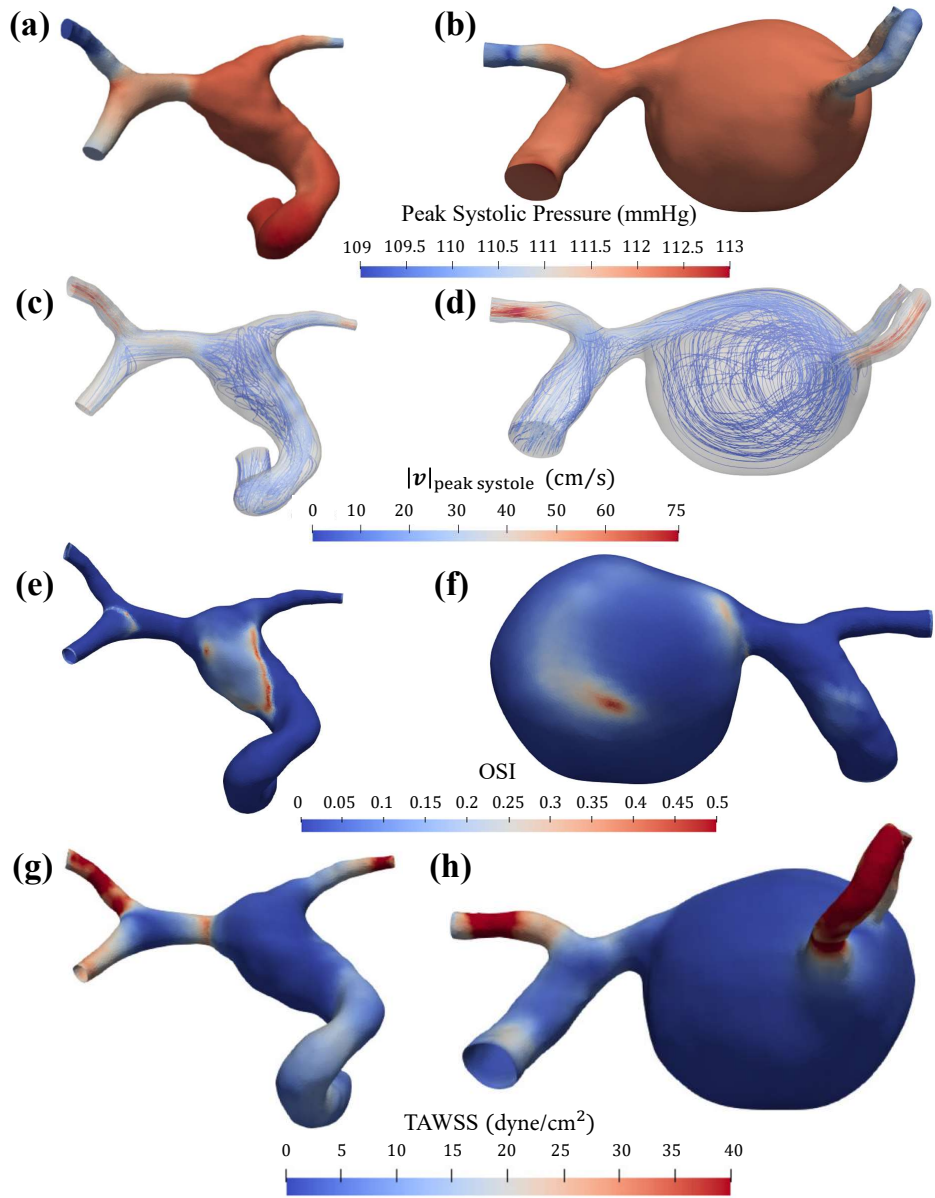}
    \caption{Hemodynamic metrics for subject S1. Each row shows a single metric: i.e.\ contours for pressure at peak systole, flow streamlines at peak systole, contours of the oscillatory shear index (OSI), and time-averaged wall shear stress magnitude. OSI was calculated using Equation \eqref{Eqn:OSI}. Panels (a), (c), (e), and (g) correspond to aneurysm S1A1, while panels (b), (d), (f), and (h) correspond to aneurysm S1A2. Abbreviations used -- $\vert\mathbf{v}\vert$: Velocity magnitude, ${\rm OSI}$: Oscillatory Shear Index, ${\rm TAWSS}$: Time-averaged wall shear stress magnitude.}
    \label{fig:S1_Hemodynamic}
\end{figure}

\begin{figure}
    \centering
    \includegraphics[width=0.75\textwidth]{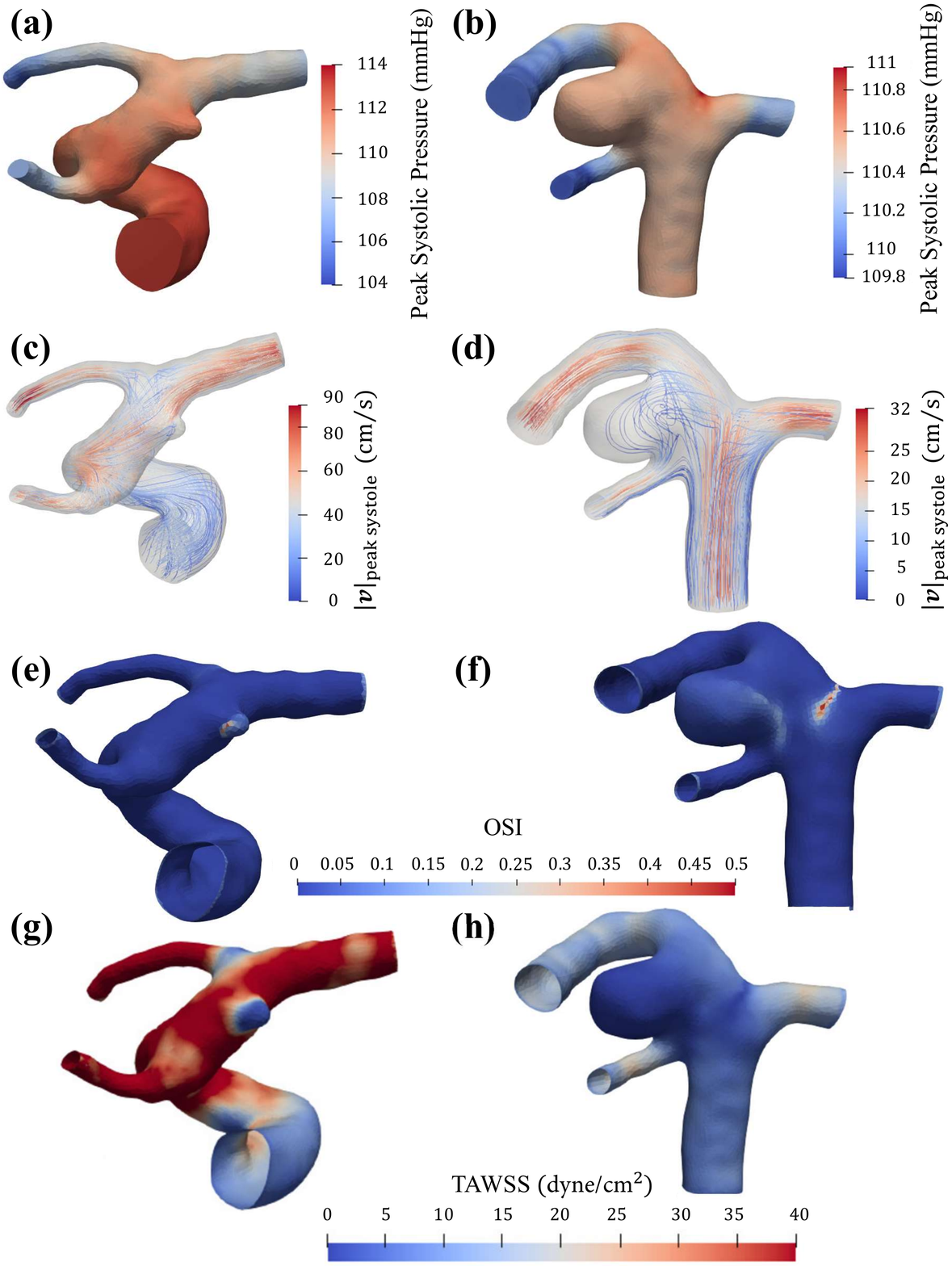}
    \caption{Hemodynamic metrics for subject S2. Each row shows a single metric: i.e.\ contours for pressure at peak systole, flow streamlines at peak systole, contours of the oscillatory shear index (OSI), and time-averaged wall shear stress magnitude. OSI was calculated using Equation \eqref{Eqn:OSI}. Panels (a), (c), (e), and (g) correspond to aneurysm S2A1, while panels (b), (d), (f), and (h) correspond to aneurysm S2A2. Abbreviations used -- $\vert\mathbf{v}\vert$: Velocity magnitude, ${\rm OSI}$: Oscillatory Shear Index. Note that the ranges of velocity and pressure values change across panels (a)-(d), as made clear by the individual color bars in each panel, ${\rm TAWSS}$: Time-averaged wall shear stress magnitude.}
    \label{fig:S2_Hemodynamic}
\end{figure}

Panels (a)-(d) in Figures \ref{fig:S1_Hemodynamic} and \ref{fig:S2_Hemodynamic} show the pressure and velocity at peak systole for the stable and growing aneurysms in each subject, respectively. The corresponding Reynolds numbers at the inlet at peak systole are reported in Table \ref{tb:Results}. 

\subsection{Hemodynamic Metrics}
\label{sec:heodynamic_metrics}

Previous hemodynamic modeling studies have focused on wall shear stress or more specifically abnormalities in wall shear stress magnitude and changes in wall shear stress direction over the cardiac cycle (given by the oscillatory shear index) as indirect measures of assessing vascular degradation and remodeling. Therefore, in this study, we also present the time-averaged wall shear stress (referred to as ${\rm TAWSS}$)  and the oscillatory shear index (OSI) data for each aneurysm.
The oscillatory shear index (OSI) is defined as 
\begin{equation}
    {\rm OSI}(\mathbf{x}) = \frac{1}{2}\left(1-\frac{\left|\frac{1}{T}\int_0^T \boldsymbol{\tau}_w(\mathbf{x},t)\,dt\right|}{\frac{1}{T}\int_0^T |\boldsymbol{\tau}_w(\mathbf{x},t)|\,dt}\right),
    \label{Eqn:OSI}
\end{equation}
where $\boldsymbol{\tau}_w(\mathbf{x},t)$ is the wall shear stress vector at a given location $\mathbf{x}$ along the wall at time instant $t$ of the cardiac cycle. OSI is indicative of the directionality of wall shear stress. OSI values close to 0 imply that the wall shear stress vector does not change direction over the cardiac cycle, whereas values close to 0.5 indicate reversal ($180^\circ$ flip) in the direction of the wall shear stress vector on a time-averaged basis. 
Previous studies suggested a correlation between regions of high OSI, indicative of high oscillatory shear stress on the aneurysmal wall, and aneurysm progression \cite{XTSM14}. Panels (e) and (f) of Figures \ref{fig:S1_Hemodynamic} and \ref{fig:S2_Hemodynamic} show contours of OSI plotted on the surface of each aneurysm. As seen from the figures, in both subjects, the stable and growing aneurysms demonstrate regions of high OSI ($\approx 0.5$). However, we do not find any association between high OSI and aneurysm growth. Furthermore, there appears to be no significant pattern of which regions are exposed to high OSI in the aneurysm geometries being considered. 

Next, in panels (g) and (h) of Figures \ref{fig:S1_Hemodynamic} and \ref{fig:S2_Hemodynamic}, we show the contours of the magnitude of time-averaged wall shear stress, ${\rm TAWSS}$, for both subjects.
In previous studies, abnormally high and low wall shear stresses have been shown to elicit inflammatory responses resulting in aneurysm initiation and growth. As reported by Meng \textit{et al.}~\cite{MTXS14}, low wall shear stress is predominantly responsible for aneurysmal growth via matrix metalloprotease (MMP) induced degradation of the extracellular matrix (ECM). Consequently, we compare the areas characterized by abnormally low time-averaged wall shear stress  for the stable and growing aneurysms. First, we defined a quantity ${\rm TAWSS}^*$ as
\begin{equation}
    {\rm TAWSS}^* (\mathbf{x}) = \frac{{\rm TAWSS}(\mathbf{x})}{\tilde{\tau}_{\rm parent}},
    \label{Eqn:TAWSS*}
\end{equation}
where ${\rm TAWSS}(\mathbf{x})$ is the time-averaged wall shear stress magnitude at a location $\mathbf{x}$ on the aneurysm (as shown in sub-panels (g) and (h) in Figures \ref{fig:S1_Hemodynamic} and \ref{fig:S2_Hemodynamic}), and $\tilde{\tau}_{parent}$ is the spatiotemporal averaged wall shear stress on the parent/feeding artery averaged over its surface as well as over a single cardiac cycle. The contour plots showing the distribution of ${\rm TAWSS}^*$ over the aneurysmal surface are shown in Appendix C of the Supplemental Material. Subsequently, based on previous studies \cite{LGRMGLGSA17}, a threshold of ${\rm TAWSS}^*<0.1$ was used to classify regions under abnormally low shear. 

Table \ref{tb:Results} lists the percentage of area under low ${\rm TAWSS}^*$ as compared to the total surface area of the aneurysm. It can be observed that abnormally low shear regions are present in both stable and growing aneurysms. However, within the same subject, the area under low ${\rm TAWSS}^*$ is larger (almost by a factor of 2 in S1 and more in S2) in the growing aneurysm, as compared to the stable aneurysm. This observation, which is consistent with observations from previous studies \cite{BRMMALHSYS08,SOTTHKMK04}, suggests that the proportion of area under low wall shear (relative to the total aneurysmal surface area) contributes to aneurysm growth or stability, rather than the mere presence of an area exposed to low wall shear.  Physiologically, aneurysm growth is driven by the imbalance between arterial wall repair and wall degradation caused by aberrant hemodynamics \cite{MTXS14,FCRA19}. Therefore, we hypothesize that the larger proportion of area under low ${\rm TAWSS}^*$ on the aneurysm leads to larger regions of the aneurysmal wall being subjected to cell degradation and apoptosis, tipping the balance towards aneurysm growth.

\begin{table}[ht]
\caption{Reynolds number at peak systole and percentage of area under abnormal biomechanical loads for each aneurysm.}
\begin{center}
\begin{tabular}{MMMM}
\toprule
Aneurysm & ${\rm Re}$ & Area under low shear$^*$ & Area under low shear and low OStI$^*$\\
\midrule
S1A1&329&18.3&-- \\
S1A2&250&34.9&23.3 \\
S2A1&490&3.1&3.3 \\
S2A2&292&58.8&32.8 \\
\bottomrule
\end{tabular}
\end{center}
\label{tb:Results}
{\noindent $^*$As a percentage of the total surface area of the aneurysm.}
\end{table}
Note that the inlet Reynolds numbers at peak systole given in Table~\ref{tb:Results} are calculated according to ${\rm Re} = {\rho_f v_\text{peak}D_{\rm vessel}}/{\mu_f}$, where $v_\text{peak}$ is the centerline velocity at peak systole, and $D_{\rm vessel}=2\sqrt{A_\text{inlet}/\pi}$ is the effective vessel diameter at the inlet plane, with $A_\text{inlet}$ being the surface area of the inlet face.

\subsection{Structural Metrics}
\label{sec:structural_metrics}

One of the contributions of our proposed FSI-simulation-based analysis is the ability to compute metrics within the arterial wall in addition to the hemodynamic parameters. Here, we consider two such biomechanical metrics: the wall displacement and the oscillatory stress index referred to as OStI.
\begin{figure}
    \centering
    \includegraphics[width=0.75\textwidth]{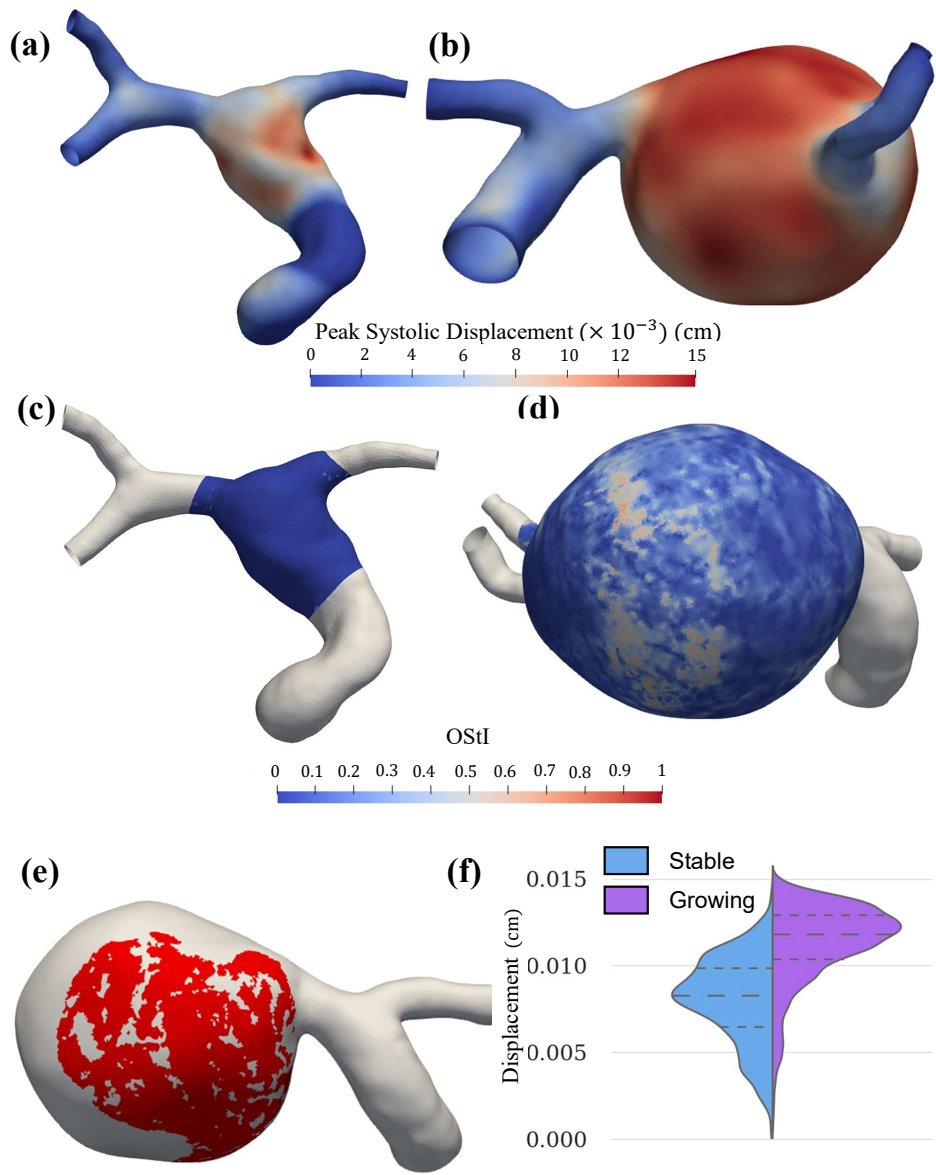}
    \caption{Biomechanical parameters for subject S1. (a) and (b) Displacement at peak systole for aneurysms S1A1 and S1A2, respectively. (c) and (d) Oscillatory stress index in aneurysms S1A1 and S1A2, respectively, calculated using Equations \eqref{Eqn:OStI} and \eqref{Eqn:Orientation}. (e) Region of overlapping low ${\rm TAWSS}^*$ ($<0.1$) and low ${\rm OStI}<0.96$) in the aneurysm S1A2. (f) Split violin plot of the distribution of peak systolic displacements (see panels (a) and (b) in the same figure) over the aneurysmal surface. The data for S1A1 is shown in the left-hand side plot, whereas the data for S1A2 is shown in the right-hand side plot. Dashed lines (lowest to highest) show the 25-50-75\% quartiles. Abbreviations used -- ${\rm TAWSS}^*$: Time-averaged wall shear stress normalized by the parent/feeding vessel spatiotemporal wall shear stress average, ${\rm OStI}$: Oscillatory Stress Index.}
    \label{fig:S1_Structural}
\end{figure}

\begin{figure}
    \centering
    \includegraphics[width=0.75\textwidth]{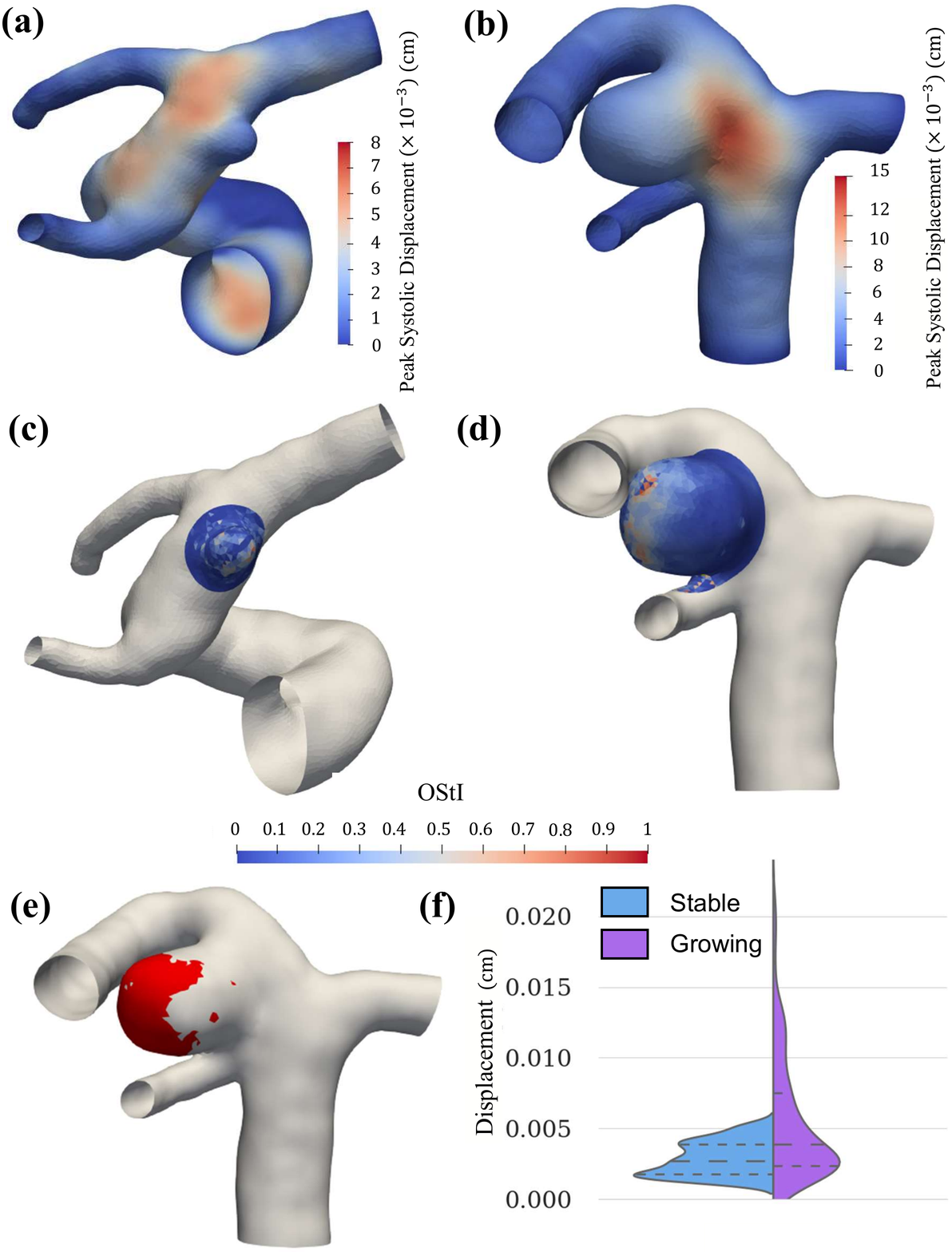}
    \caption{Biomechanical parameters for subject S2. (a) and (b) Displacement at peak systole for aneurysms S2A1 and S2A2, respectively. (c) and (d) Oscillatory stress index in aneurysms S2A1 and S2A2, respectively, calculated using Equations \eqref{Eqn:OStI} and \eqref{Eqn:Orientation}. (e) Region of overlapping low ${\rm TAWSS}^*$ ($<0.1$) and low ${\rm OStI}<0.96$) in the aneurysm S2A2. (f) Split violin plot of the distribution of peak systolic displacements (see panels (a) and (b) in the same figure) over the aneurysmal surface. The data for S2A1 is shown in the left-hand side plot,  whereas the data for S2A2 is shown in the right-hand side plot. Dashed lines (lowest to highest) show the 25-50-75\% quartiles. Abbreviations used -- ${\rm TAWSS}^*$: Time-averaged wall shear stress normalized by the parent/feeding vessel spatiotemporal wall shear stress average, ${\rm OStI}$: Oscillatory Stress Index. Note that the ranges of velocity and pressure values change across panels (a) and (b), as made clear by the individual color bars in each panel.}
    \label{fig:S2_Structural}
\end{figure}

Panels (a) and (b) of Figures \ref{fig:S1_Structural} and \ref{fig:S2_Structural} show the displacement at peak systole for each aneurysm. The corresponding spatial distribution are shown in panel (f) of Figures \ref{fig:S1_Structural} and \ref{fig:S2_Structural}, as a split violin plot. 

For both subjects, the median peak displacement is approximately 1.5 times higher in the growing aneurysms as compared to the corresponding stable aneurysms. Furthermore, while the largest displacements on the stable and growing aneurysms are not significantly different for S1, they differ by almost a factor of three for S2. Previous work has shown the impact of arterial wall displacement on aneurysm progression and eventual rupture. Vanrosomme \textit{et al.}~\cite{VETCZ15} hypothesized that increased distension of aneurysms is associated with an increased risk of rupture due to loss of mural cells. Other studies (e.g.\ Liu \textit{et al.}~\cite{LSZLQASLSLQYZ18} and the references therein) indicated that cyclic mechanical stretch leads to mural cell degradation, thus driving aneurysm growth. The findings from our study suggest that, as compared to the reference state at diastole, growing aneurysms undergo larger displacements within the cardiac cycle and therefore may be more prone to mural cell degradation as compared to their stable counterparts.

We define the oscillatory stress index (OStI) at each location as a quantity related to the time-averaged change in orientation of the maximum absolute principal stress over the cardiac cycle with respect to the orientation at diastole as reference. Mathematically, this definition is be expressed as
\begin{equation}
    {\rm OStI}(\mathbf{x}) = 1 - \langle\cos\theta\rangle(\mathbf{x}),
    \label{Eqn:OStI}
\end{equation}
where 
\begin{equation}
    \langle\cos\theta\rangle(\mathbf{x}) = \frac{1}{T}\int_0^T \cos(\theta(\mathbf{x},t))\,dt,
    \label{Eqn:Theta_avg}
\end{equation}
$T$ is the time period of the cardiac cycle, diastole is assumed to be at $t=0$, and
\begin{equation}
    \cos(\theta (\mathbf{x},t)) = \frac{|\mathbf{PD}(\mathbf{x},t)\cdot \mathbf{PD}(\mathbf{x},0)|}{|\mathbf{PD}(\mathbf{x},t)|\cdot|\mathbf{PD}(\mathbf{x},0)|}
    \label{Eqn:Orientation}
\end{equation}
is the angle between the orientation of the maximum absolute principal stress $\mathbf{PD}(\mathbf{x},t)$ at a given location $\mathbf{x}$ at time $t$ and the orientation of the maximum absolute principal stress at diastole $\mathbf{PD}(\mathbf{x},0)$ (i.e.\ $\mathbf{PD}$ at the same location $\mathbf{x}$ and $t=0$). The maximum absolute principal stress at any given location $\mathbf{x}$ and time point $t$ is the largest absolute eigenvalue of the local stress tensor. Subsequently, the orientation of this maximum absolute principal stress is the orientation of the associated eigenvector. 

Based on the definition of OStI, higher values indicate that the direction of the maximum absolute principal stress fluctuates through the cardiac cycle and lower values (close to zero) indicate that the maximum absolute principal stress orientation remains unchanged. The methodology used to compute the principal stresses and the corresponding directions is similar to that used in \cite{BHZWLKBI10}. The stress tensor on the outer surface of the arterial wall is expressed in terms of the local coordinate system comprised of two local orthogonal surface tangent vectors (given by $\mathbf{\mathfrak{t}_1}$ and  $\mathbf{\mathfrak{t}_2}$) and the local normal $\mathbf{n}$. The corresponding rotation tensor $\mathbf{R}$ can be written as:
$\begin{bmatrix} \mathbf{\mathfrak{t}}_1 & \mathbf{\mathfrak{t}}_2 & \mathbf{n} \end{bmatrix}$. Subsequently, the traction condition on the outer wall is enforced strongly by modifying the last row and column of the rotated stress tensor. However, unlike the simulations in \cite{BHZWLKBI10}, for which a zero traction condition was imposed, we prescribe a value of $-k\boldsymbol{\delta}$, where $k$ is the tissue support parameter and $\boldsymbol{\delta}$ is the displacement. In the local coordinates, the normal vector can simply be written as $\mathbf{n} = \begin{bmatrix} 0 & 0 & 1 \end{bmatrix}^\top$. Therefore, imposing $\boldsymbol{\sigma}\cdot\mathbf{n} = -k\boldsymbol{\delta}$ is equivalent to setting the elements in the last row (and column) of the rotated stress tensor equal to $-k$ times the corresponding component of the displacement expressed in the local coordinate system. The displacement expressed in the local coordinate system, denoted $\tilde{\boldsymbol{\delta}}$, and the displacement in the global coordinate system, denoted $\boldsymbol{\delta}$, are related as $\tilde{\boldsymbol{\delta}} = \mathbf{R}^\top\boldsymbol{\delta}$. 

Panels (c) and (d) of Figures \ref{fig:S1_Structural} and \ref{fig:S2_Structural} show the contour plots of OStI for the stable and growing aneurysms in both subjects. Collagen is the main constituent within the extra-cellular matrix (ECM) of the arterial wall that maintains structural integrity through dynamic cross-linking of fibers over the cardiac cycle in response to tensile stresses elicited by blood pressure \cite{MTXS14,H08}. Consequently, collagen fibers in regions of low OStI values are exposed to oscillating tensile stresses requiring repeated realignment and re-orientation of fibers over the cardiac cycle \cite{HDGH07}. We hypothesize that this type of `structural insult' is the principal driver of collagen remodeling. In combination with a compromised ECM due to MMP-induced degradation caused by low wall shear, the result is a degraded and remodeled collagen fiber network. This would reduce the capability of the arterial wall to sustain arterial stress under normal loading conditions, driving aneurysmal wall remodeling and growth. 

Subsequently, we extracted regions of overlapping low wall shear stress (i.e.\ where ${\rm TAWSS}^*<0.1$) and regions of high OStI (i.e.\ where $\langle\cos\theta\rangle < \cos(\theta_0)$) and compared the proportion of these regions between the stable and growing aneurysms in the same subject. The angle threshold $\theta_0$ for classifying regions of high OStI was set to $15^\circ$. This threshold value has been chosen such that from the point of view of using dot product computations to gauge the colinearity of the unit principal stress directions vectors on a time-averaged basis. Any values of $\langle\cos\theta\rangle$ ranging from 1 to 0.96 i.e.\ less than 5\% error margin are considered colinear. The corresponding overlapping area (in red online, or darker shade in print) is shown in panel (e) of Figures \ref{fig:S1_Structural} and \ref{fig:S2_Structural}. We observe that while such regions occupy a sizeable proportion of the growing aneurysms, they are almost either entirely absent or negligibly small ($<5\%$) in stable aneurysms (see Table \ref{tb:Results}). This observation indicates that oscillatory arterial wall stresses, combined with regions of low wall shear stresses, may be potential biomarkers for aneurysm progression.

\subsection{Limitations}
\label{sec:Limitations}

The main limitations of the present analysis lie in the modeling assumptions made for the arterial wall and boundary conditions for the FSI simulations. Since the models were based on retrospective imaging data, obtaining subject-specific flow and pressure measurements for these aneurysms was not possible. Therefore, flow boundary conditions used in the analysis, i.e.\ the RCR outflow boundary conditions and the inlet flow waveforms, were not subject-specific. However, as explained in Sections \ref{sec:heodynamic_metrics} and \ref{sec:structural_metrics}, the flow and structural metrics obtained from our simulations correspond to a physiologically realizable cardiovascular state in these subjects. Therefore, the relative comparisons between stable and growing aneurysms in the same subject are meaningful. 

The arterial wall thickness model is also not subject-specific. This limitation arises due to a lack of vessel wall information in standard-of-care imaging, including CTA, MRA and X-ray angiography data. This information is vital in obtaining patient-specific estimates of the biomechanical parameters, in particular the OStI and vessel wall displacement. With additional cases of subjects with multiple unruptured aneurysms showing different progression statuses, a sensitivity analysis could be performed to determine the wall thickness variations on the reported biomechanical parameters. However, obtaining subject-specific wall thickness distributions, by incorporating advanced imaging modalities, such as black-blood MRI or amplified MRI \cite{HMBCHL20,TNGRZYMKH18}, would ultimately lead to a more accurate estimation of arterial stresses and displacements. This topic is also left for future work.

The assumption that the arterial wall can be modeled as an isotropic neo-Hookean material is a simplification of the physiological wall response. Moreover, the elasticity model we used does not account for changes in material properties due to the aneurysmal disease. These limitations arose primarily due to a lack of patient-specific wall material properties. While this precludes us from making any patient-specific predictions for our subject cases, future studies following the methodology in the represent analysis, incorporating subject-specific variations in arterial wall properties or constitutive models such as the Holzapfel- Gasser-Ogden (HGO) model \cite{HDGH07}, could enable accounting for the multi-component nature, anisotropy, and local variation in material properties of the arterial wall, thereby allowing for subject-specific predictions of the reported biomechanical parameters.

Blood vessels in the brain are in contact with various tissues, which, in general, have a nonlinear and anisotropic spatiotemporal response. While we account for heterogeneity in tissue support, depending on the nature of the contact, the svFSI platform currently lacks the capability to model the nonlinear elastic or time-dependent behavior of the surrounding tissue. With regard to the spring-dashpot boundary condition used, our analysis neglects viscous damping due to the presence of CSF (cerebrospinal fluid) in the subarachnoid spaces. With advancements in the computational framework, and by incorporating subject-specific wall motion data, it may become possible to fine-tune tissue support parameters. The present analysis could be extended along these lines in the future.

Lastly, in this study, only two subjects were considered, owing to the rarity of subjects with multiple, unruptured and untreated aneurysms who, in addition to possessing both growing and stable aneurysms, have undergone longitudinal imaging studies. Performing any assessment of the reported biomarkers, with statistical significance/power needed for translation of this approach into clinical assessments and management of cerebral aneurysms, will require a larger cohort of subjects with multiple aneurysms for whom longitudinal imaging data will be obtained. This topic is left for future work with prospectively recruited subjects.

\section{Conclusion}
\label{sec:Conc}

We modeled coupled blood flow and vessel wall motion and displacement in two subjects with multiple cerebral aneurysms. Using standard-of-care patient imaging data, we developed physiologically realistic computational models, which account for both the nonuniform aneurysmal wall thickness and the effect of surrounding tissue support on the outer wall's motion and displacement. Hemodynamic and structural metrics were computed and compared between stable and growing aneurysms in the same subject. Specifically, the time-averaged wall shear stress, the oscillatory shear index, and peak systolic displacement were considered. Additionally, we defined and computed a novel metric, the oscillatory stress index, which is indicative of the fluctuations in the orientation of the largest arterial principal stress. 

Significant differences were observed between stable and growing aneurysms in the same subject in the area under low shear, peak systolic displacement, and the area under combined low shear and low oscillatory stress index. More importantly, the proportion of aneurysmal area exposed to both low shear and oscillating arterial stresses was large ($\approx 23$ and $33\%$ in S1 and S2, respectively) in the growing aneurysms, compared to the corresponding stable aneurysms, for which such areas were either nonexistent or less than 5\% of the total area. Based on these initial results, we hypothesize that the presence of significant regions under this abnormal combined loading, which signifies a large degree of degradation and remodeling of collagen in the arterial wall, may be used as a potential predictor of aneurysmal growth. 

The proposed computational framework, and the associated biomechanical factors, provide a proof-of-concept for a novel approach to quantitative assessment of the risk of growth in aneurysms. Thus, our study sets the groundwork for future studies on biomechanical risk factors in aneurysmal disease in order to help clinicians with the risk stratification of  cerebral aneurysms.

\section*{Acknowledgements}
This research was supported by an award from the Kati Lorge Chair of Research of the Brain Aneurysm Foundation. We further acknowledge the resources of Extreme Science and Engineering Discovery Environment (XSEDE) and the Texas Advanced Computing Center (TACC), supported by the National Science Foundation under grant number ACI-1548562 (project MCH200007). Simulations were also performed on the community clusters at the Rosen Center for Advanced Computing (RCAC) at Purdue University.

\bibliographystyle{elsarticle-num} 

\bibliography{references}
\clearpage

\begin{appendices}

\begin{center}
{\LARGE SUPPLEMENTARY MATERIAL FOR BIO-22-1154}\\[5mm]
{\LARGE Comparative Assessment of Biomechanical Parameters in Subjects With Multiple Cerebral Aneurysms Using Fluid--Structure Interaction Simulations}\\[3mm]
{\large Tanmay C.\ Shidhore$^{1}$,
Aaron A.\ Cohen-Gadol$^{2}$,
Vitaliy L.\ Rayz$^{1,3,\dagger}$, and
Ivan C.\ Christov$^{1,\dagger}$}\\[1mm]
{\small\it $^{1}$School of Mechanical Engineering, Purdue University, West Lafayette, Indiana 47907}\\
{\small\it $^{2}$Department of Neurological Surgery, Indiana University School of Medicine, Indianapolis, Indiana 46202}\\
{\small\it $^{3}$Weldon School of Biomedical Engineering, Purdue University, West Lafayette, Indiana 47907}\\[1mm]
{\small $\dagger$ Corresponding authors; \email{vrayz@purdue.edu}, \email{christov@purdue.edu}.}
\end{center}

\setcounter{page}{1}
\setcounter{equation}{0}
\setcounter{figure}{0}
\numberwithin{equation}{section}
\renewcommand{\thefigure}{\thesection\arabic{figure}}
\numberwithin{figure}{section}

\section{Grid-Independence Study}

\begin{figure}[ht]
    \centering
    \includegraphics[width=0.75\textwidth]{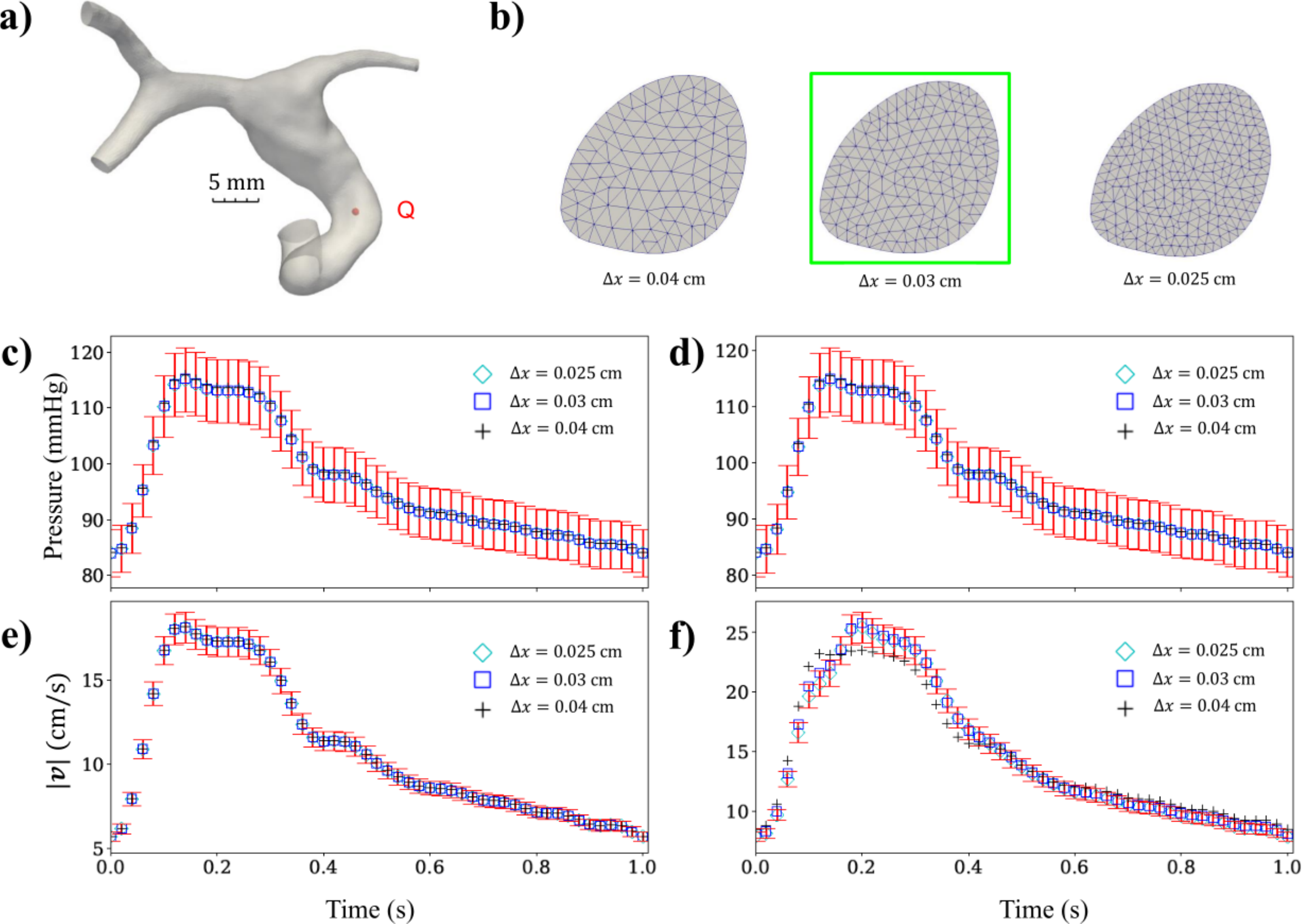}
    \caption{Pressure and velocity data over a cardiac cycle at the inlet plane (panels (c) and (e)) and at a point (panels (d) and (f)) in the interior of the feeding vessel (shown in (a)) for different mesh resolutions (shown in (b)). The error bars of each plot point show a deviation of 5\% from the corresponding value on the finest mesh ($\Delta x = 0.025~\si{\centi\meter}$). Abbreviations used -- $\vert\mathbf{v}\vert$: Velocity magnitude}
    \label{fig:GS_core}
\end{figure}

To ensure that computational quantities reported, such as pressure, velocity, wall shear stress, and those derived thereof, were independent of the grid resolution used in the simulations, a grid sensitivity analysis for the fluid domain's mesh was performed. The types of meshes, properties, and refinement levels used are reported in Table \ref{tb:Grid_optimization}. A two-step approach was used to establish the grid independence of the simulations. First, a core mesh resolution was determined for which the pressure and velocity averaged over the inlet face, as well as at a point in the interior of the parent vessel (see Figure \ref{fig:GS_core}(a)), were independent of this core mesh's resolution. Second, varying degrees of mesh refinement was implemented via boundary layers close to the fluid-solid interface, on top of the chosen core mesh resolution from the previous step. This second step ensured that the computed wall shear stress was independent of the near-wall mesh resolution. 

Figure~\ref{fig:GS_core} shows the chosen representative pressure and velocity magnitude plotted over a single cardiac cycle. From these plots, we observe that the pressure and velocity values computed on the mesh with $\Delta x = 0.03~\si{\centi\meter}$ lie within a 5\% of the values computed on the most refined mesh ($\Delta x = 0.025~\si{\centi\meter}$). Therefore, a core mesh resolution of $\Delta x = 0.03~\si{\centi\meter}$ was chosen as the optimal core mesh resolution for our simulations.

\begin{table}[ht]
\caption{Details of meshes used for the grid-independence study.}
\begin{center}
\begin{tabular}{llllll}
\toprule
$\Delta x_{\rm core}$ (cm) & $N_{\rm BL}$ & $N_{\rm elements}$ & $N_{\rm nodes}$ & Max.\ ${\rm CFL}$ \\
\midrule
0.04 &&169,370&32,743&0.59\\[2mm]
\multirow{4}{*}{0.03}&0&396,208&73,797&0.6 \\
&3&400,225&74,673&4.7\\
&4&400,565&74,820&4.48\\
&5&401,517&75,076&4.61\\[2mm]
0.025 &&681,305&124,604&0.88 \\[2mm]
\bottomrule 
\end{tabular}
\end{center}
\label{tb:Grid_optimization}
\end{table}

Based on the plots in Figure \ref{fig:GS_core}, we observed that the pressure and velocity magnitude values computed on both the coarse and medium meshes (i.e.\ those with $\Delta x = 0.04~\si{\centi\meter}$ and $\Delta x = 0.03~\si{\centi\meter}$) lie within a 5\% margin of the values computed on the fine grid. However, in Figure \ref{fig:GS_core}f, the velocity magnitude for the coarse grid ($\Delta x = 0.04~\si{\centi\meter}$) lies beyond this tolerance margin. Therefore, $\Delta x = 0.03~\si{\centi\meter}$ was determined to be the core mesh resolution of choice. 

\begin{figure}
    \centering
    \includegraphics[width=0.75\textwidth]{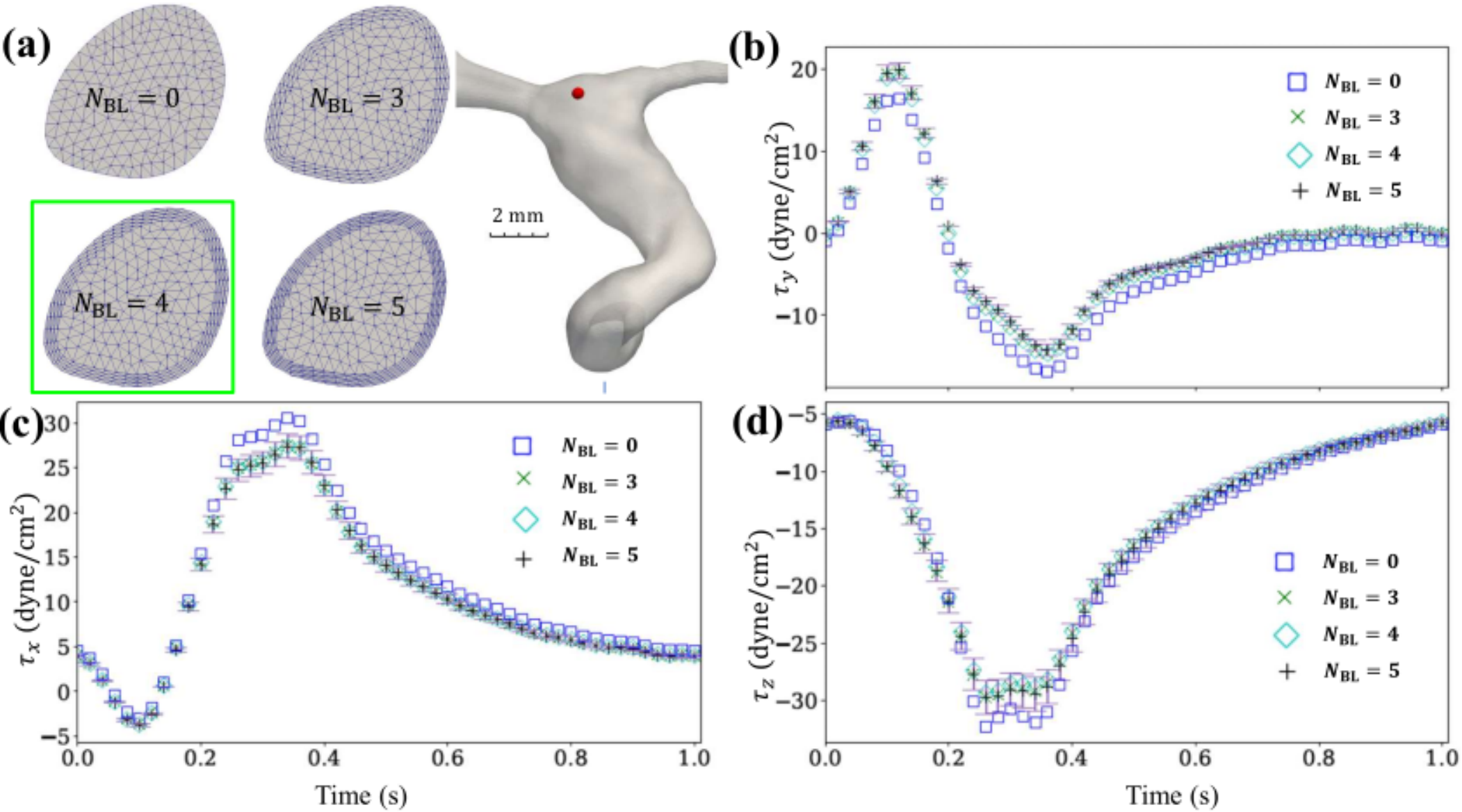}
    \caption{Components (shown in panels (b), (c) and (d)) of the WSS $\tau$ over a cardiac cycle at a point on the aneurysmal surface (shown by a dot in the model geometry in (a)), for different numbers of boundary layers each. $N_{\rm BL}$ represents the number of layers of boundary layer elements. Here, $N_{\rm BL} = 0$ represents a mesh without boundary layer refinement. The error bars on each plot point show a deviation of 5\% from the corresponding value on the mesh with the largest number of boundary layer refinements (i.e.\ $N_{\rm BL} = 5$).}
\label{fig:GS_BL}
\end{figure}

Next, Figure \ref{fig:GS_BL} shows the $x$, $y$, and $z$ components of the WSS (wall shear stress) computed at a point on the surface of the ascending aorta. Here, the core mesh resolution was identical in all cases ($\Delta x = 0.03~\si{\centi\meter}$). However, close to the fluid-solid interface, different numbers of layers of mesh refinement (0, 3, 4, and 5) were considered (see Figure \ref{fig:GS_core}(a)). From Figure~\ref{fig:GS_BL}(b)-(d), we observed a nontrivial difference ($> 5\%$) between the surface shear stress values computed on meshes with and without mesh refinement. Furthermore, meshes with different levels of mesh refinement ($N_{\rm BL}=3,4$, and $5$) yield shear stress values within the above tolerance limit with minor differences in the computational time. Therefore, we proceeded with a mesh refinement level of $N_{\rm BL} = 4$, to balance the need for increased resolution with the corresponding computational cost.

A constant time step of $\Delta t=10^{-4}~\si{s}$ was used for all cases. Table \ref{tb:Grid_optimization} reports an estimate of the maximum cell-based Courant number computed for each of the meshes used, over a single cardiac cycle. The cell-based Courant number is defined as ${\rm CFL} = {\vert\mathbf{v}\vert \Delta t}/{\Delta x}$, where $\vert\mathbf{v}\vert$ is the velocity magnitude at the cell center, $\Delta t$ is the time step size, and $\Delta x$ is a length scale computed for each cell as $\Delta x = \mathcal{V}^{1/3}$, where $\mathcal{V}$ is the cell volume.

We observed that only a few elements ($<1\%$ of the total number of elements) that were close to the outlet faces exceeded the threshold of having a maximum ${\rm CFL} > 1$. This observation, together with the fact that the time-integration scheme implemented in svFSI is implicit  \cite{CH93}, allowed us to use the same time step size of $\Delta t=10^{-4}~\si{\sec}$ for the subsequent FSI simulations as well.

\section{Orientation of Principal Stresses at End Diastole}

\begin{figure}
    \centering
    \includegraphics[width=0.75\textwidth]{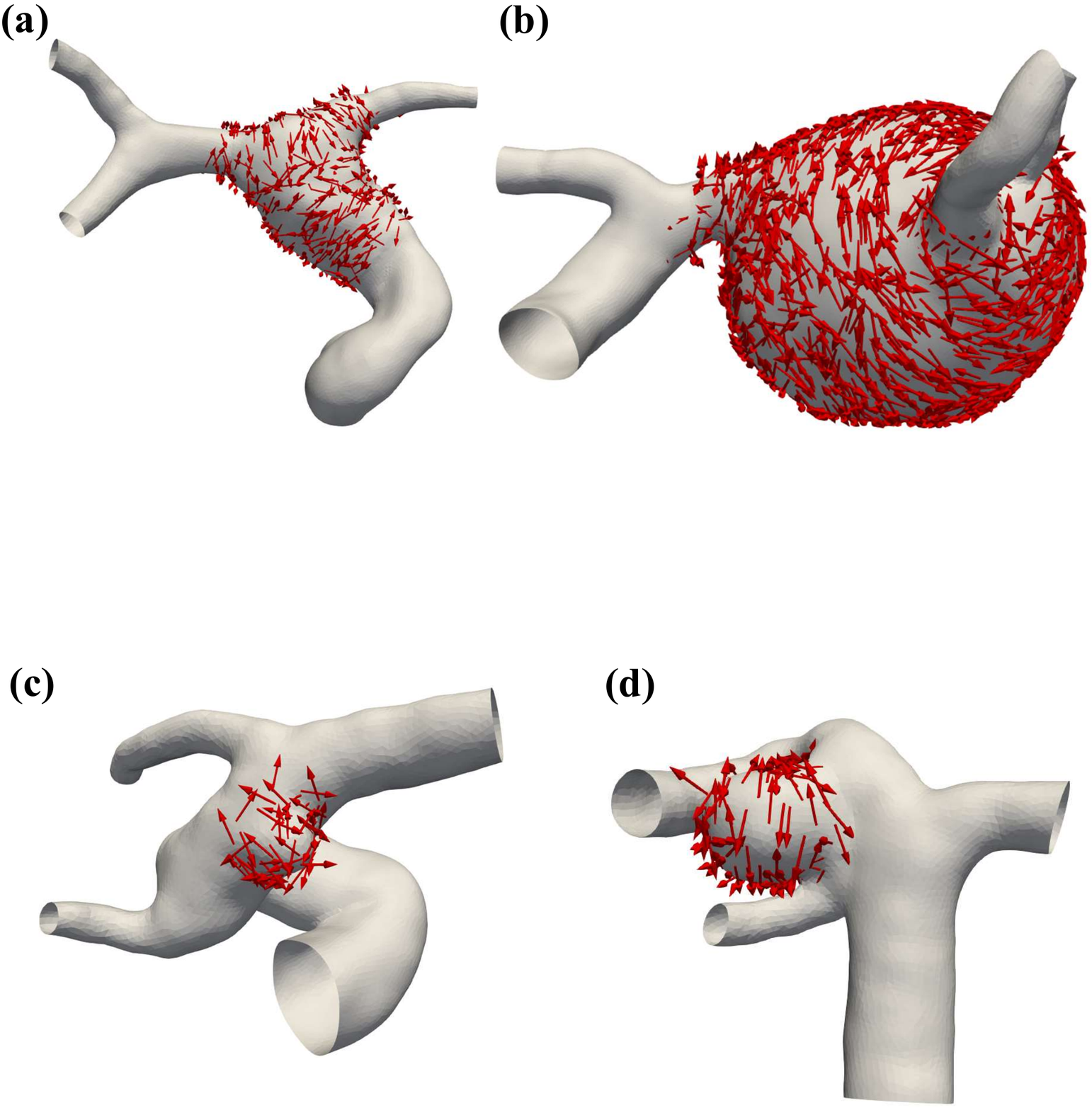}
    \caption{Orientation (red arrows) of the largest principal stress direction at end-diastole in Subject 1 (top row) and Subject 2 (bottom row). Panels a) and c) show the stable aneurysms and panels b) and d) show the growing aneurysms.}
\label{fig:PS_Orientation}
\end{figure}

To ensure completeness of the OStI data discussed in Section \ref{sec:Results}, the orientation of the largest principal stress at end-diastole is shown in Figure \ref{fig:PS_Orientation}. It can be seen that the direction of the largest principal stress is tangential to the aneurysmal surface. This observation, together with the data reported in Figures \ref{fig:S1_Structural} and \ref{fig:S2_Structural}, 
provides a means for visualizing the extent of oscillations of the direction of the largest principal stress on a cardiac-cycle average-basis.

\section{Normalized Hemodynamic and Structural Metrics}

\begin{figure}
    \centering
    \includegraphics[width=0.75\textwidth]{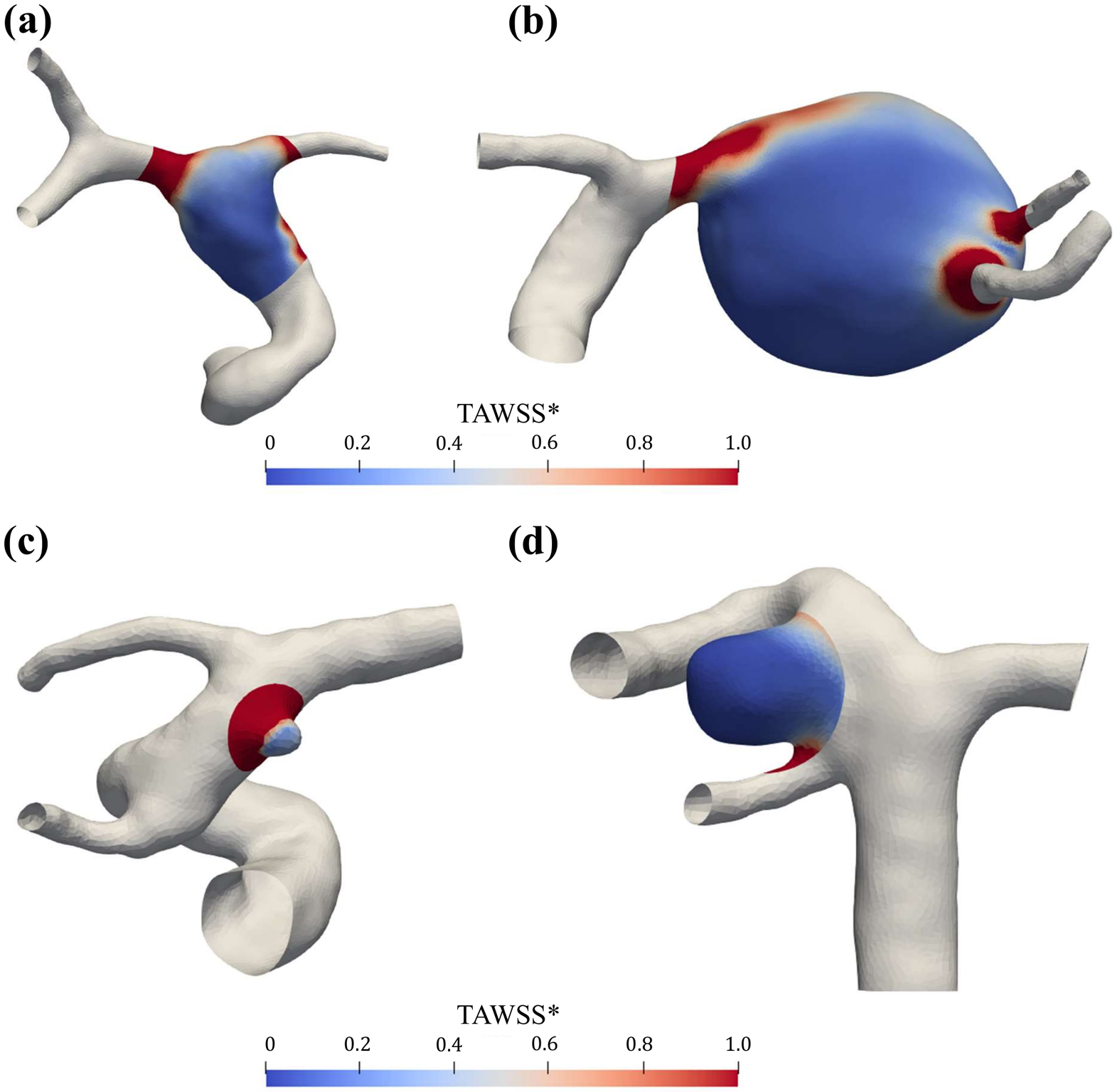}
    \caption{Contour plots of ${\rm TAWSS}^*$, as per Equation \eqref{Eqn:TAWSS*} over the aneurysmal surface for (a) S1A1 (b) S1A2 (c) S2A1 (d) S2A2.}
\label{fig:Norm_metrics}
\end{figure}

\begin{figure}
    \centering
    \includegraphics[width=0.75\textwidth]{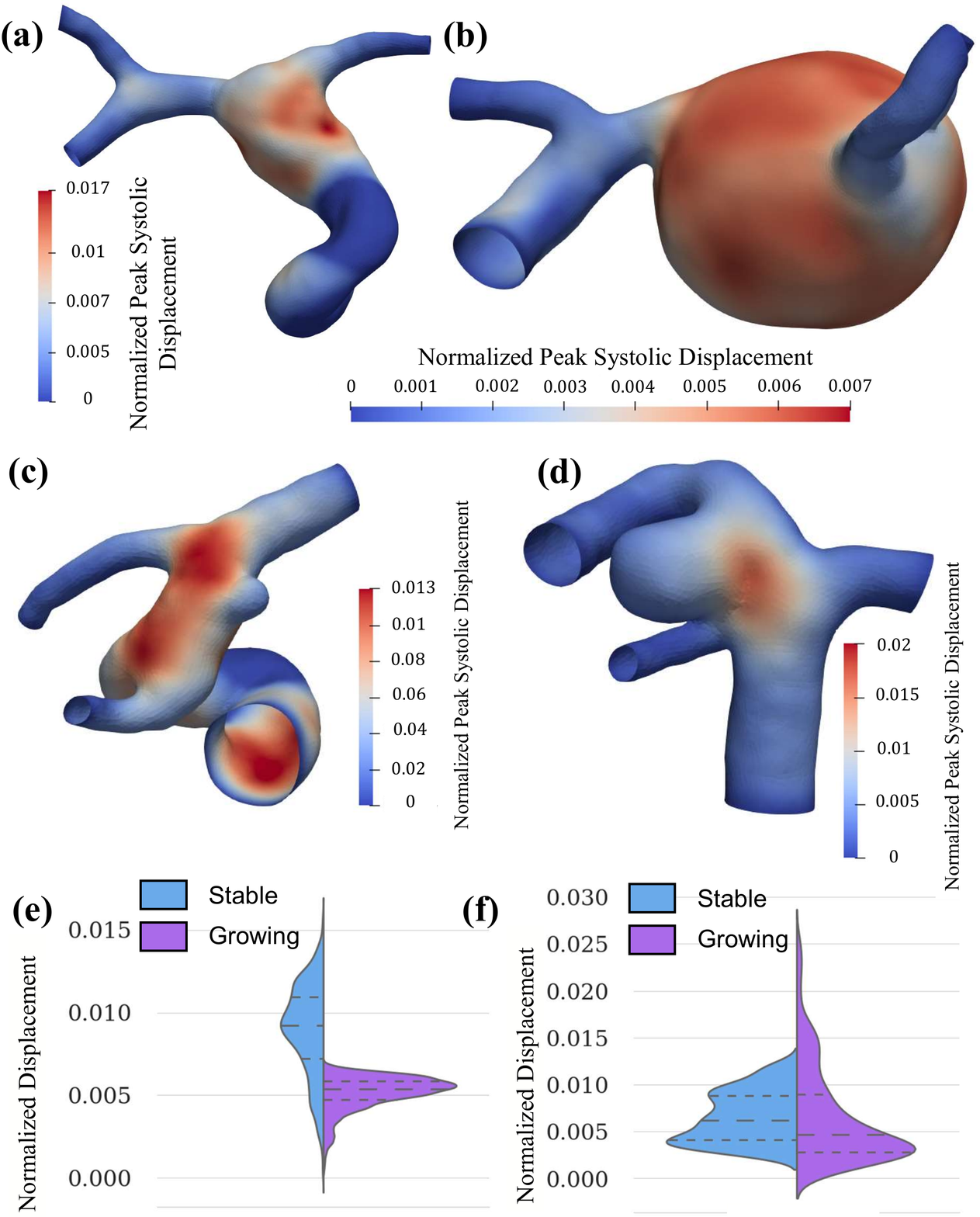}
    \caption{(a) and (b) Normalized displacement at peak systole for aneurysms S1A1 and S1A2, respectively. (c) and (d) Normalized displacement at peak systole for aneurysms S1A1 and S1A2, respectively. (e) and (f) Split violin plot of the distribution of peak systolic displacements in Subjects 1 and 2 respectively (see panels (a), (b), (c), and (d) in the same figure) over the aneurysmal surface. The data for the stable is shown in the left-hand side plot, whereas the data for the growing aneurysm is shown in the right-hand side plot in each sub-panel. Dashed lines (lowest to highest) show the 25-50-75\% quartiles.}
    \label{fig:Structural_Normalized_Displacement}
\end{figure}

This appendix reports data on the corresponding normalized metrics for TAWSS and the peak systolic displacement from Section \ref{sec:Results}. 
Figure \ref{fig:Norm_metrics} shows the contour plots of ${\rm TAWSS}^*$, which are used to classify regions of low shear on the aneurysmal surface.
Figure \ref{fig:Structural_Normalized_Displacement} shows the wall displacement data at peak systole, normalized by the initial aneurysm diameters reported in Table \ref{tb:Subject_characteristics}. 
In contrast to the unnormalized deformation data shown in Figures \ref{fig:S1_Structural} and \ref{fig:S2_Structural}, a comparison of the normalized data shows no clear pattern  across the two subject cases. The median normalized deformation is found to be approximately 2 and 1.5 times higher in the stable aneurysms as compared to the growing aneurysms in subjects 1 and 2, respectively. Furthermore, while the maximum peak normalized displacement for the growing aneurysm in subject 2 is twice the corresponding value in the stable aneurysm, the ratio reverses for subject 1.
\end{appendices}

\end{document}